\documentclass[superscriptaddress,onecolumn,pre]{revtex4}

\usepackage{amsmath}
\usepackage{graphicx,color}

\newcommand{\be}{\begin{equation}}
\newcommand{\ee}{\end{equation}}
\newcommand{\bea}{\begin{eqnarray}}
\newcommand{\eea}{\end{eqnarray}}

\begin{document}
\sloppy


\title{Kinetic theory of spatially homogeneous systems with long-range interactions: \\
I. General results}

\author{Pierre-Henri Chavanis}
\affiliation{Laboratoire de Physique Th\'eorique (IRSAMC), CNRS and UPS, Universit\'e de Toulouse, F-31062 Toulouse, France}

\begin{abstract}

We review and complete the existing literature on the kinetic theory
of spatially homogeneous systems with long-range interactions taking
collective effects into account. The evolution of the system as a
whole is described by the Lenard-Balescu equation which is valid in a
weak coupling approximation. When collective effects are neglected it
reduces to the Landau equation and when collisions (correlations) are
neglected it reduces to the Vlasov equation. The relaxation of a test
particle in a bath is described by a Fokker-Planck equation involving
a diffusion term and a friction term. For a thermal bath, the
diffusion and friction coefficients are connected by an Einstein
relation. General expressions of the diffusion and friction
coefficients are given, depending on the potential of interaction and
on the dimension of space. We also discuss the scaling with $N$
(number of particles) or with $\Lambda$ (plasma parameter) of the
relaxation time towards statistical equilibrium. Finally, we consider
the effect of an external stochastic forcing on the evolution of the system and
compute the corresponding term in the kinetic equation.

\end{abstract}

\maketitle

\section{Introduction}
\label{sec_introduction}

Recently, the dynamics and thermodynamics of systems with long-range interactions has been a subject of considerable interest in statistical mechanics \cite{houches,assise,oxford,cdr}. Some of these systems, perhaps the most important, have been studied for a long time. This is the case of Coulombian plasmas, self-gravitating systems and two-dimensional vortices. In the last few years, other systems with long-range interactions have been introduced (or revived) and studied. They include one and two dimensional plasmas, one and two dimensional self-gravitating systems and toy models such as the Hamiltonian mean field (HMF) model  and  the one dimensional ring model \cite{cdr}.  Although these models are relatively academic, they are of great conceptual interest to better understand the peculiar properties of long-range interacting (LRI) systems. Furthermore, they are  simpler to simulate numerically and this is certainly valuable to perform very accurate comparisons between numerical experiments and theory.  Finally, these toy models may find unexpected applications in physics. For example, the HMF model presents many features analogous to the Colson-Bonifacio Hamiltonian model describing single-pass free electron lasers (FEL).

The equilibrium statistical mechanics of systems with long-range interactions is now relatively well understood \cite{cdr}. One important feature of these systems is that they may exhibit ensemble inequivalence and negative specific heats first observed in astrophysics (see \cite{katz,paddy,ijmpb} for reviews).  The out-of-equilibrium properties of these systems are also very interesting. In particular, long-range interacting systems may be stuck in non-Boltzmannian quasi stationary states (QSS) that persist for a very long time before finally reaching the Boltzmann distribution. These QSSs are stable steady states of the Vlasov equation which governs the ``collisionless'' evolution of the system. In order to understand the different timescales involved in the dynamics, it is important to develop a kinetic theory of systems with long-range interactions. Since these systems are generically spatially inhomogeneous, the kinetic theory is relatively complicated and requires the introduction of angle-action variables \cite{angleaction,kindetail,heyvaerts,newangleaction}. However, in some cases (e.g. plasma physics), a stable homogeneous phase exists, and, in other cases (e.g. stellar dynamics), it is possible to make a local approximation and treat the system as if it were spatially homogeneous. If we restrict ourselves to spatially homogeneous systems, most of the work has already been done in the context of plasma physics and stellar dynamics since the kinetic equations that govern the ``collisional'' evolution of a spatially homogeneous plasma or a stellar system, namely the Landau, Lenard-Balescu and Fokker-Planck equations, can immediately be transposed to other systems with long-range interactions.  However, in the details, the diffusion and friction coefficients, and the relaxation time, sensitively depend on the form of the potential of interaction and on the dimension of space.  It is therefore useful to develop a general formalism common to all systems with long-range interactions and, from this general formalism, describe specific models.

Since the kinetic equations initially introduced in plasma physics and
stellar dynamics are now studied by a larger community, it may be of
interest to regroup in a self-consistent paper the derivation of the
basic equations (Vlasov, Landau, Lenard-Balescu and Fokker-Planck) to
the attention of a wider audience. This is done in this paper (paper
I) where we extend the derivation given in \cite{cdr} to a more
general class of potentials of interaction in arbitrary dimensions of
space. For the unity of the subject, we follow their presentation as closely as possible. We also consider the effect of an external stochastic forcing on the
evolution of the system and compute the corresponding term in the
kinetic equation. In the second paper (Paper II), we use these general 
results to determine the explicit expressions of the diffusion
and friction coefficients, and the relaxation time, for several
systems with long-range interactions. Our two papers are therefore
essentially pedagogical and aim to synthesize and develop the kinetic
theory of systems with long-range interactions. They also show what is
common to all systems with long-range interactions and what is
specific to such or such system.

\section{Evolution of the system as a whole}
\label{sec_whole}

\subsection{Klimontovich and Vlasov equations}
\label{sec_kv}

Let us consider a system of material particles in interaction in a
$d$-dimensional space. Their dynamics is described by the Hamilton
equations
\begin{eqnarray}
m\frac{d{\bf r}_i}{dt}=\frac{\partial H}{\partial {\bf v}_i},\qquad m \frac{d{\bf v}_i}{dt}=-\frac{\partial H}{\partial {\bf r}_i},
\label{pvm1}
\end{eqnarray}
with the Hamiltonian
\begin{eqnarray}
H=\sum_{i=1}^{N} \frac{1}{2}mv_i^2+m^2\sum_{i<j}u(|{\bf r}_i-{\bf r}_j|),
\label{pvm2}
\end{eqnarray}
where $u(|{\bf r}-{\bf r}'|)$ is the potential of interaction and $m$ the individual mass of the particles. For an isolated system, the total energy $E=H$ is conserved. We assume that the potential of interaction decays at large distances as $r^{-\gamma}$ with $\gamma\le d$.  In that case, the potential  is said to be long-ranged \cite{cdr}. Long-range potentials include the gravitational and the Coulombian potentials, corresponding to $\gamma=d-2$.

We introduce the discrete distribution function  $f_d({\bf r},{\bf v},t)=m\sum_i\delta({\bf r}-{\bf r}_i(t))\delta({\bf v}-{\bf v}_i(t))$. Differentiating this expression with respect to time and using the equations of motion (\ref{pvm1})-(\ref{pvm2}), we find that $f_d({\bf r},{\bf v},t)$ satisfies an equation of the form
\begin{eqnarray}
\frac{\partial f_d}{\partial t}+{\bf v}\cdot \frac{\partial f_d}{\partial {\bf r}}-\nabla\Phi_d\cdot \frac{\partial f_d}{\partial {\bf v}}=0,
\label{pvm9}
\end{eqnarray}
where
\begin{eqnarray}
\Phi_d({\bf r},t)=\int u(|{\bf r}-{\bf r}'|)f_d({\bf r}',{\bf v}',t)\, d{\bf r}'d{\bf v}',
\label{pvm9bis}
\end{eqnarray}
is the discrete  potential produced by $f_d({\bf r},{\bf v},t)$.
This equation is {\it exact} and contains the same information as the Hamiltonian system (\ref{pvm1})-(\ref{pvm2}). In plasma physics, it is called the Klimontovich equation  \cite{klimontovich}.

We now introduce a smooth distribution function $f({\bf r},{\bf v},t)=\langle f_{d}({\bf r},{\bf v},t)\rangle$ corresponding to an average of $f_{d}({\bf r},{\bf v},t)$ over a large number of initial conditions. We then write $f_d=f+\delta f$ where $\delta f$ denotes the fluctuations around the smooth distribution. Similarly, we write $\Phi_d=\Phi+\delta\Phi$. Substituting these decompositions in Eq. (\ref{pvm9}), we obtain
\begin{equation}
\frac{\partial f}{\partial t}+\frac{\partial\delta f}{\partial t}+{\bf v}\cdot\frac{\partial f}{\partial {\bf r}}+{\bf v}\cdot\frac{\partial \delta f}{\partial {\bf r}}-\nabla\Phi\cdot \frac{\partial f}{\partial {\bf v}}-\nabla\Phi\cdot \frac{\partial \delta f}{\partial {\bf v}}-\nabla \delta\Phi\cdot\frac{\partial f}{\partial {\bf v}}-\nabla\delta\Phi\cdot \frac{\partial \delta f}{\partial {\bf v}}=0.
\label{pvm10}
\end{equation}
Taking the average of this equation over the initial conditions, we get
\begin{equation}
\frac{\partial f}{\partial t}+{\bf v}\cdot\frac{\partial f}{\partial {\bf r}}-\nabla\Phi\cdot \frac{\partial f}{\partial {\bf v}}=\frac{\partial}{\partial {\bf v}}\cdot \left\langle \delta f\nabla\delta\Phi\right\rangle,
\label{pvm11}
\end{equation}
where the right hand side can be interpreted as a ``collision'' term. It does not really correspond to direct collisions between particles but rather  to close encounters, or correlations, due to finite $N$ effects. Subtracting this expression from Eq. (\ref{pvm10}), we obtain an equation for the perturbation
\begin{equation}
\frac{\partial \delta f}{\partial t}+{\bf v}\cdot\frac{\partial \delta f}{\partial {\bf r}}-\nabla\Phi\cdot \frac{\partial \delta f}{\partial {\bf v}}-\nabla\delta\Phi\cdot \frac{\partial f}{\partial {\bf v}}=\frac{\partial}{\partial {\bf v}}\cdot  (\delta f\nabla\delta\Phi)-\frac{\partial}{\partial {\bf v}}\cdot \left\langle \delta f\nabla\delta\Phi\right\rangle.
\label{pvm12}
\end{equation}
These equations are still exact since no approximation has been made so far.

Using dimensional analysis in Eq. (\ref{pvm2}), we find that the kinetic energy and the potential energy are comparable if $Nmv^2\sim N^2 m^2 u$. Therefore, the energy, the temperature and the dynamical time scale like $E\sim N^2m^2u$, $\beta^{-1}\sim mv^2\sim Nm^2u$ and $t_D\sim R/v\sim R/\sqrt{Nmu}$ (where $R\sim V^{1/d}$ is the system size). We now consider the thermodynamic limit
 $N\rightarrow +\infty$ in such a way that the normalized energy $\epsilon=E/(N^2m^2 u)$, the normalized
temperature $\eta=\beta Nm^2u$ and the normalized dynamical time $\tau_D\sim t_D\sqrt{Nmu}/R$ are of order unity. By a suitable normalization of the physical quantities, it corresponds to $N\rightarrow +\infty$ with $V\sim 1$, $m\sim 1$, $E\sim N$, $\beta\sim 1$, $t_D\sim 1$, and $u\sim 1/N$.  This is equivalent to putting $1/N$ in front of the potential energy in the Hamiltonian (Kac prescription) \cite{kac}. Since the strength of the potential goes to zero as $N\rightarrow +\infty$, we are dealing with {\it weak} long-range potentials. In this scaling, the energy is extensive and the temperature intensive. Furthermore, $f/N\sim 1$, $\Phi\sim 1$, $\delta f/N\sim 1/\sqrt{N}$ and $\delta\Phi\sim 1/\sqrt{N}$. We can therefore consider an expansion of the equations of the problem in terms of the small parameter $1/N$ which is a  measure of the ``graininess'' of the distribution (as $1/N$ approaches zero, collisional effects disappear) \footnote{In plasma physics, the system is infinite. In that case, the size $R$ is replaced by the Debye length $\lambda_D$, the dynamical time $t_D$ by the inverse of the plasma pulsation $\omega_P^{-1}$, and the small parameter $1/N$  by the graininess parameter $g=1/\Lambda$, where the plasma parameter $\Lambda=n\lambda_D^d$ gives the number of electrons in the Debye sphere \cite{klimontovich}. In plasma physics, the $1/N$ expansion is replaced by a $1/\Lambda$ expansion.}. Considering Eq. (\ref{pvm11}), we see that the collision term (r.h.s.) scales like $1/N$ as compared to the advective term (l.h.s.). Therefore, for $N\rightarrow +\infty$, Eq. (\ref{pvm11}) reduces to the Vlasov equation
\begin{equation}
\frac{\partial f}{\partial t}+{\bf v}\cdot\frac{\partial f}{\partial {\bf r}}-\nabla\Phi\cdot\frac{\partial f}{\partial {\bf v}}=0,\qquad \Phi({\bf r},t)=\int u(|{\bf r}-{\bf r}'|)f({\bf r}',{\bf v}',t)\, d{\bf r}'d{\bf v}'.
\label{pvm13}
\end{equation}
The Vlasov equation is obtained when ``collisions'' (more properly
granular effects, discreteness effects or finite $N$ corrections)
between particles are neglected. The Vlasov equation can also be
obtained from the Liouville equation, writing the BBGKY hierarchy and
neglecting correlations between particles. This is valid, for a fixed
interval of time, when $N \rightarrow +\infty$. In that case, the mean
field approximation becomes exact and the $N$-body distribution
function can be factorized in a product of $N$ one-body distributions,
resulting in Eq. (\ref{pvm13}). The ``Vlasov equation'', which is also
called the collisionless Boltzmann equation (CBE), was introduced by
Jeans \cite{jeans} in stellar dynamics and by Vlasov \cite{vlasov} in
plasma physics. In fact, it is valid for a large class of systems with
long-range interactions \cite{cdr}. The exactness of the Vlasov
equation when $N\rightarrow +\infty$ was proven rigorously by Braun \&
Hepp \cite{bh} for non singular potentials.

Starting from an unsteady or unstable initial condition, the Vlasov
equation, coupled to a long-range potential of interaction, is known
to develop a complicated mixing process leading to the formation of a
quasistationary state (QSS) on a coarse-grained scale. This process,
which is purely collisionless and driven by mean field effects, is
known as violent relaxation.  It takes place on a very short
timescale, of the order of a few dynamical times $t_D\sim 1$. A
statistical theory of the Vlasov equation has been developed by
Lynden-Bell \cite{lb} in astrophysics to predict the QSS that results
from violent relaxation, assuming ergodicity (or, at least, efficient
mixing). A similar statistical theory has been developed by Miller
\cite{miller} and Robert \& Sommeria \cite{rs} for the Euler equation
in two-dimensional (2D) hydrodynamics.  The process of violent
relaxation towards a non-Boltzmannian QSS has also been extensively
studied for the HMF model \cite{cdr}. In some cases, violent
relaxation is {\it incomplete} due to lack of ergodicity or efficient
mixing \cite{incomplete}.

On longer timescales, ``collisions'' (more properly correlations) between particles develop and the system deviates from the Vlasov dynamics. For $t\rightarrow +\infty$, we expect that the system will reach a statistical equilibrium state. In the thermodynamic limit $N \rightarrow +\infty$, the mean field approximation is exact \cite{ms}. The statistical equilibrium state in the microcanonical ensemble is obtained by maximizing the Boltzmann entropy $S=-\int (f/m)\ln(f/m)\, d{\bf r}d{\bf v}$ at fixed mass $M=\int f\, d{\bf r}d{\bf v}$ and energy $E=(1/2)\int f v^2\, d{\bf r}d{\bf v}+(1/2)\int\rho\Phi\, d{\bf r}$ where $\rho({\bf r},t)=\int f({\bf r},{\bf v},t)\, d{\bf v}$ is the mass density. This variational principle determines the most probable macrostate $f({\bf r},{\bf v})$. Writing the first order variations as $\delta S-\beta\delta E-\alpha\delta M=0$ where $\beta$ and $\alpha$ are Lagrange multipliers, we obtain the mean field Boltzmann distribution \footnote{Equation (\ref{pvm14}) determines all the critical points of constrained entropy, i.e. those that cancel the first order variations. Of course, among these solutions, only entropy {\it maxima} are physically relevant. We must therefore consider the sign of the second order variations of entropy and discard entropy minima and saddle points. If several entropy maxima are found for the same values of the constraints, we must distinguish fully stable states (global maxima) and metastable states (local maxima). }:
\begin{equation}
f({\bf r},{\bf v})=Ae^{-\beta m (\frac{v^2}{2}+\Phi({\bf r}))},\qquad \Phi({\bf r})=\int u(|{\bf r}-{\bf r}'|)\rho({\bf r}')\, d{\bf r}'.
\label{pvm14}
\end{equation}
For the Newtonian interaction (gravity), the equilibrium potential is solution of the Boltzmann-Poisson equation. In the case of plasmas with a neutralizing background, the statistical equilibrium state is spatially homogeneous and Eq. (\ref{pvm14}) reduces to the Maxwell distribution. We must remember, however, that the statistical theory is based on an assumption of ergodicity (or efficient mixing) and on the postulate that all the accessible microstates are equiprobable. There is no guarantee that this is always true for systems with long-range interactions. In order to prove that the system reaches the Boltzmann distribution (\ref{pvm14}), and in order to determine the relaxation time (in particular its scaling with the number $N$ of particles), we must develop a kinetic theory of systems with long-range interactions \cite{cdr,bgm,kindetail}.

\subsection{Landau and Lenard-Balescu equations}
\label{sec_lb}

As we have previously indicated, the ``collisions'' between particles can be neglected for times shorter than $N t_D$ where $t_D$ is the dynamical time. If we want to describe the  evolution of the system on a longer timescale, we must take finite $N$ corrections into account. At the level $1/N$, we can neglect the quadratic term on the right hand side of Eq. (\ref{pvm12}). Indeed, the left hand side is of order $1/\sqrt{N}$ and the right hand side is of order $1/N\ll 1/\sqrt{N}$. We therefore obtain the set of coupled equations
\begin{equation}
\frac{\partial f}{\partial t}+{\bf v}\cdot \frac{\partial f}{\partial {\bf r}}-\nabla\Phi\cdot \frac{\partial f}{\partial {\bf v}}=\frac{\partial}{\partial {\bf v}}\cdot \left\langle \delta f \nabla\delta\Phi\right\rangle,
\label{lb1}
\end{equation}
\begin{equation}
\frac{\partial\delta f}{\partial t}+{\bf v}\cdot \frac{\partial \delta f}{\partial {\bf r}}-\nabla\delta\Phi\cdot \frac{\partial f}{\partial {\bf v}}-\nabla\Phi\cdot \frac{\partial \delta f}{\partial {\bf v}}=0.
\label{lb2}
\end{equation}
They form the starting point of the quasilinear theory which is valid in a weak coupling approximation \cite{pitaevskii}. It can be shown that these equations describe the evolution of the system under the effect of two-body collisions (higher order correlations are neglected). If we restrict ourselves to spatially homogeneous distributions, the field $-\nabla\Phi$ vanishes. In that case, Eqs. (\ref{lb1})-(\ref{lb2}) reduce to the coupled equations
\begin{equation}
\frac{\partial f}{\partial t}=\frac{\partial}{\partial {\bf v}}\cdot \left\langle \delta f \nabla\delta\Phi\right\rangle,
\label{lb3}
\end{equation}
\begin{equation}
\frac{\partial\delta f}{\partial t}+{\bf v}\cdot \frac{\partial \delta f}{\partial {\bf r}}-\nabla\delta\Phi\cdot \frac{\partial f}{\partial {\bf v}}=0.
\label{lb4}
\end{equation}
These equations are valid as long as the spatially homogeneous distribution function is Vlasov stable so that it evolves under the sole effect of collisions. We shall assume that the fluctuations evolve rapidly compared to the transport time scale, so that time variation of $f$ and $\Phi$  can be neglected in the calculation of the collision term (Bogoliubov ansatz). Therefore, for the purpose of solving Eq. (\ref{lb4}) and obtaining the correlation function $\left\langle \delta f \delta\Phi\right\rangle$, we shall regard $f({\bf v})$ as constant in time. Actually, the time dependence of $f$ is given by Eq. (\ref{lb3}).  Only the fluctuation term $\langle \delta f\delta\Phi\rangle$ drives $f$ and then its evolution is much slower than the evolution of $\delta f$. Indeed, $f$ changes on a timescale of the order $Nt_D$ (or larger) while  $\left\langle \delta f \delta\Phi\right\rangle$ relaxes, by Landau damping, to its asymptotic form on a much shorter timescale, of the order of the dynamical time $t_D$. After the correlation function has been obtained as a functional of $f$, the time dependence of $f$ may be reinserted in the kinetic equation. This is an adiabatic hypothesis which is valid for $N\gg 1$. With this approximation, Eqs. (\ref{lb3})-(\ref{lb4}) can be solved with the aid of Fourier-Laplace transforms and the collision term can be explicitly calculated. This is the approach  presented in \cite{pitaevskii} to derive the Lenard-Balescu equation.

The Fourier-Laplace transform of the fluctuations  of the distribution function  $\delta f$ is defined by
\begin{equation}
\delta \tilde f({\bf k},{\bf v},\omega)=\int \frac{d{\bf r}}{(2\pi)^d}\int_{0}^{+\infty}dt\, e^{-i({\bf k}\cdot{\bf r}-\omega t)}\delta f({\bf r},{\bf v},t).
\label{lb5}
\end{equation}
This expression for the Laplace transform is valid for ${\rm Im}(\omega)$ sufficiently large. For the remaining part of the complex $\omega$ plane, it is defined by an analytic continuation. The inverse transform is
\begin{equation}
\delta f({\bf r},{\bf v},t)=\int d{\bf k}\int_{\cal C}\frac{d\omega}{2\pi}\, e^{i({\bf k}\cdot{\bf r}-\omega t)}\delta\tilde f({\bf k},{\bf v},\omega),
\label{lb6}
\end{equation}
where the Laplace contour ${\cal C}$ in the complex $\omega$ plane must pass above all poles of the integrand. Similar expressions hold for the fluctuations  of the potential $\delta\Phi$. We note that, for periodic potentials, the integral over ${\bf k}$ is replaced by a discrete summation over the different modes. If we take the  Fourier-Laplace transform of Eq. (\ref{lb4}), we find that
\begin{equation}
-\delta\hat{f}({\bf k},{\bf v},0)-i\omega\, \delta\tilde f({\bf k},{\bf v},\omega)+i {\bf k}\cdot{\bf v}\, \delta\tilde f({\bf k},{\bf v},\omega)-i {\bf k}\cdot\frac{\partial f}{\partial {\bf v}}\, \delta\tilde\Phi({\bf k},\omega)=0,
\label{lb7}
\end{equation}
where the first term is the spatial Fourier transform of the initial value
\begin{equation}
\delta\hat{f}({\bf k},{\bf v},0)=\int\frac{d{\bf r}}{(2\pi)^d}\, e^{-i{\bf k}\cdot {\bf r}}\delta f({\bf r},{\bf v},0).
\label{lb8}
\end{equation}
The foregoing equation can be rewritten
\begin{equation}
\delta\tilde f ({\bf k},{\bf v},\omega)=\frac{{\bf k}\cdot \frac{\partial f}{\partial {\bf v}}}{{\bf k}\cdot {\bf v}-\omega}\delta\tilde\Phi({\bf k},\omega)+\frac{\delta\hat f({\bf k},{\bf v},0)}{i({\bf k}\cdot {\bf v}-\omega)},
\label{lb9}
\end{equation}
where the first term on the right hand side corresponds to ``collective effects'' and the second term is related to the initial condition. The fluctuations of the potential are related to the fluctuations of the distribution function by a convolution
\begin{equation}
\delta\Phi({\bf r},t)=\int u(|{\bf r}-{\bf r}'|)\delta f({\bf r}',{\bf v}',t)\, d{\bf r}'d{\bf v}'.
\label{deltaconvol}
\end{equation}
Taking the Fourier-Laplace transform of this equation, we obtain
\begin{equation}
\delta\tilde\Phi({\bf k},\omega)=(2\pi)^d\hat{u}(k)\int\delta\tilde f({\bf k},{\bf v},\omega)\, d{\bf v}.
\label{lb10}
\end{equation}
Substituting Eq. (\ref{lb9}) in Eq. (\ref{lb10}), we find that the Fourier-Laplace transform of the fluctuations of the potential is related to the initial condition by
\begin{equation}
\delta\tilde\Phi({\bf k},\omega)=(2\pi)^d\frac{\hat{u}(k)}{\epsilon({\bf k},\omega)}\int d{\bf v}\, \frac{\delta\hat f({\bf k},{\bf v},0)}{i({\bf k}\cdot {\bf v}-\omega)},
\label{lb11}
\end{equation}
where the dielectric function is defined by
\begin{equation}
\epsilon({\bf k},\omega)=1-(2\pi)^d\hat{u}(k)\int  \frac{{\bf k}\cdot \frac{\partial f}{\partial {\bf v}}}{{\bf k}\cdot {\bf v}-\omega}\, d{\bf v}.
\label{lb12}
\end{equation}
The Fourier-Laplace transform of the fluctuations of the distribution function is
then given by Eq. (\ref{lb9}) with Eq. (\ref{lb11}). In these equations, and in the following,
$f$ stands for $f({\bf v})$ and $f'$ stands for $f({\bf v}')$. The dispersion relation associated with the linearized Vlasov equation corresponds to $\epsilon({\bf k},\omega)=0$ (see Appendix \ref{sec_disp}). If the system is Vlasov stable, $\epsilon({\bf k},\omega)$ does not vanish for any real $\omega$, so Eq. (\ref{lb11}) is well-defined. If collective
effects were neglected in Eq. (\ref{lb9}), we would obtain Eq. (\ref{lb11}) with $\epsilon({\bf k},\omega)=1$. This shows that, due to collective effects, the bare potential of interaction $\hat{u}(k)$ is replaced by a ``dressed'' potential $\hat{u}(k)/\epsilon({\bf k},\omega)$ taking into account the polarization of the medium.

We can use these expressions to compute the collision term appearing on the right hand side of Eq. (\ref{lb3}). One has
\begin{equation}
\left\langle\delta f \nabla\delta\Phi\right\rangle=\int d{\bf k}\int_{\cal C} \frac{d\omega}{2\pi}\int d{\bf k'}\int_{\cal C} \frac{d\omega'}{2\pi} \, i {\bf k}' e^{i({\bf k}\cdot {\bf r}-\omega t)}e^{i({\bf k}'\cdot {\bf r}-\omega' t)}\langle \delta\tilde f({\bf k},{\bf v},\omega)\delta\tilde\Phi({\bf k}',\omega')\rangle.
\label{lb13}
\end{equation}
Using Eq. (\ref{lb9}), we find that
\begin{equation}
\langle \delta\tilde f({\bf k},{\bf v},\omega)\delta\tilde\Phi({\bf k}',\omega')\rangle=\frac{{\bf k}\cdot \frac{\partial f}{\partial {\bf v}}}{{\bf k}\cdot {\bf v}-\omega}\langle \delta\tilde\Phi({\bf k},\omega)\delta\tilde\Phi({\bf k}',\omega')\rangle+\frac{\langle \delta\hat f({\bf k},{\bf v},0)\delta\tilde\Phi({\bf k}',\omega')\rangle}{i({\bf k}\cdot {\bf v}-\omega)}.
\label{lb14}
\end{equation}
The first term corresponds to the self-correlation of the potential, while the second term corresponds to the correlations between the fluctuations of the potential and the fluctuations of  the distribution function at time $t=0$. Let us consider these two terms separately.

From Eq. (\ref{lb11}), we obtain
\begin{equation}
\langle \delta\tilde\Phi({\bf k},\omega)\delta\tilde\Phi({\bf k}',\omega')\rangle=-(2\pi)^{2d}\frac{\hat{u}(k)\hat{u}(k')}{\epsilon({\bf k},\omega)\epsilon({\bf k}',\omega')}\int d{\bf v}d{\bf v}'\, \frac{\langle \delta\hat f({\bf k},{\bf v},0)\delta\hat f({\bf k}',{\bf v}',0)\rangle}{({\bf k}\cdot {\bf v}-\omega)({\bf k}'\cdot {\bf v}'-\omega')}.
\label{lb15}
\end{equation}
Using the expression of the auto-correlation of the fluctuations at $t=0$ given by (see Appendix \ref{sec_auto}):
\begin{equation}
\langle \delta\hat f({\bf k},{\bf v},0)\delta\hat f({\bf k}',{\bf v}',0)\rangle=\frac{1}{(2\pi)^d}\delta({\bf k}+{\bf k}')\delta({\bf v}-{\bf v}') m f({\bf v}),
\label{lb16}
\end{equation}
we find that
\begin{equation}
\langle \delta\tilde\Phi({\bf k},\omega)\delta\tilde\Phi({\bf k}',\omega')\rangle=(2\pi)^{d}m\frac{\hat{u}(k)^2}{\epsilon({\bf k},\omega)\epsilon(-{\bf k},\omega')}\delta({\bf k}+{\bf k}')\int d{\bf v}\, \frac{f({\bf v})}{({\bf k}\cdot {\bf v}-\omega)({\bf k}\cdot {\bf v}+\omega')}.
\label{lb17}
\end{equation}
Considering only the contributions that do not decay in time, it can be shown \cite{pitaevskii} that $\lbrack ({\bf k}\cdot {\bf v}-\omega)({\bf k}\cdot {\bf v}+\omega')\rbrack^{-1}$ can be substituted by $(2\pi)^2\delta(\omega+\omega')\delta({\bf k}\cdot {\bf v}-\omega)$. Then, using the property $\epsilon(-{\bf k},-\omega)=\epsilon({\bf k},\omega)^*$, one finds that the correlations of the fluctuations of the potential are given by
\begin{equation}
\langle \delta\tilde\Phi({\bf k},\omega)\delta\tilde\Phi({\bf k}',\omega')\rangle=(2\pi)^{d+2}m\frac{\hat{u}(k)^2}{|\epsilon({\bf k},\omega)|^2}\delta({\bf k}+{\bf k}')\delta(\omega+\omega')\int  \delta({\bf k}\cdot {\bf v}-\omega)  f({\bf v})\, d{\bf v}.
\label{lb18}
\end{equation}

Similarly, one finds that the second term on the right hand side of Eq. (\ref{lb14}) is given by
\begin{equation}
\frac{\langle \delta\hat f({\bf k},{\bf v},0)\delta\tilde\Phi({\bf k}',\omega')\rangle}{i({\bf k}\cdot {\bf v}-\omega)}=(2\pi)^{2}m\frac{\hat{u}(k')}{\epsilon({\bf k}',\omega')}\delta({\bf k}+{\bf k}')\delta(\omega+\omega')   \delta({\bf k}\cdot {\bf v}-\omega)f({\bf v}).
\label{lb19}
\end{equation}

From Eq. (\ref{lb18}), we get the contribution to Eq. (\ref{lb13}) of the first term of Eq. (\ref{lb14}). It is given by 
\begin{equation}
\left\langle\delta f \nabla\delta\Phi\right\rangle_i^{I}=-i (2\pi)^{d+1} m \int d{\bf k}\int_{\cal C}\frac{d\omega}{2\pi}\int  d{\bf v}'  \, {k}_i \frac{{\bf k}\cdot \frac{\partial f}{\partial {\bf v}}}{{\bf k}\cdot {\bf v}-\omega} \frac{\hat{u}(k)^2}{|\epsilon({\bf k},\omega)|^2}f({\bf v}')\delta({\bf k}\cdot {\bf v}'-\omega).
\label{lb20}
\end{equation}
Using the Landau prescription $\omega\rightarrow \omega+i0^+$ and the Plemelj formula,
\begin{equation}
\frac{1}{x\pm i 0^+}={\cal P}\left (\frac{1}{x}\right )\mp i\pi\delta(x),
\label{lb21}
\end{equation}
where ${\cal P}$ denotes the principal value, we can replace $1/({\bf k}\cdot {\bf v}-\omega-i0^+)$ by $+i\pi\delta({\bf k}\cdot {\bf v}-\omega)$. Then, integrating on $\omega$,
we obtain
\begin{equation}
\left\langle\delta f \nabla\delta\Phi\right\rangle_i^{I}=\pi (2\pi)^{d} m \int d{\bf k}\, d{\bf v}'  \, k_ik_j   \frac{\hat{u}(k)^2}{|\epsilon({\bf k},{\bf k}\cdot {\bf v})|^2}\delta\lbrack {\bf k}\cdot ({\bf v}-{\bf v}')\rbrack f({\bf v}')\frac{\partial f}{\partial v_j}({\bf v}).
\label{lb23}
\end{equation}

From Eq. (\ref{lb19}), we get the contribution to Eq. (\ref{lb13}) of the second term of Eq. (\ref{lb14}). It is given by
\begin{equation}
\left\langle\delta f \nabla\delta\Phi\right\rangle_i^{II}= m \int d{\bf k} \, {k}_i \frac{\hat{u}(k)}{|\epsilon({\bf k},{\bf k}\cdot {\bf v})|^2}{\rm Im}\, \epsilon({\bf k},{\bf k}\cdot {\bf v})  f({\bf v}).
\label{lb24}
\end{equation}
Using the Landau prescription $\omega\rightarrow \omega+i0^+$ and the Plemelj formula (\ref{lb21}), the imaginary part of the dielectric function (\ref{lb12}) is
\begin{equation}
{\rm Im}\, \epsilon({\bf k},\omega)=-\pi(2\pi)^d\hat{u}(k)\int {\bf k}\cdot\frac{\partial f}{\partial {\bf v}}\delta({\bf k}\cdot {\bf v}-\omega)\,  d{\bf v}.
\label{lb25}
\end{equation} Substituting this expression in Eq. (\ref{lb24}), we obtain
\begin{equation}
\left\langle\delta f \nabla\delta\Phi\right\rangle_i^{II}=-\pi (2\pi)^{d} m \int d{\bf k}\, d{\bf v}'  \, k_ik_j   \frac{\hat{u}(k)^2}{|\epsilon({\bf k},{\bf k}\cdot {\bf v})|^2}\delta\lbrack {\bf k}\cdot ({\bf v}-{\bf v}')\rbrack f({\bf v})\frac{\partial f}{\partial {v'}_j}({\bf v}').
\label{lb26}
\end{equation}
Finally, regrouping Eqs. (\ref{lb3}), (\ref{lb23}) and (\ref{lb26}), we end up on the kinetic equation
\begin{equation}
\frac{\partial f}{\partial t}=\pi (2\pi)^{d}m\frac{\partial}{\partial v_i}  \int d{\bf k} \, d{\bf v}'  \, k_ik_j  \frac{\hat{u}(k)^2}{|\epsilon({\bf k},{\bf k}\cdot {\bf v})|^2}\delta\lbrack {\bf k}\cdot ({\bf v}-{\bf v}')\rbrack\left (\frac{\partial}{\partial {v}_{j}}-\frac{\partial}{\partial {v'}_{j}}\right )f({\bf v},t)f({\bf v}',t).
\label{lb27}
\end{equation}
This equation, taking  collective effects into account, is the Lenard-Balescu equation of plasma physics. It was originally derived in a very different manner \cite{lenard,balescu2} and, in fact, several derivations exist in the literature. 

If we neglect collective effects and take $|\epsilon({\bf k},{\bf k}\cdot {\bf v})|=1$, Eq. (\ref{lb27}) becomes
\begin{equation}
\frac{\partial f}{\partial t}=\pi (2\pi)^{d}m\frac{\partial}{\partial v_i}  \int d{\bf k} \, d{\bf v}'  \, k_ik_j  \hat{u}(k)^2\delta\lbrack {\bf k}\cdot ({\bf v}-{\bf v}')\rbrack\left (\frac{\partial}{\partial {v}_{j}}-\frac{\partial}{\partial {v'}_{j}}\right )f({\bf v},t)f({\bf v}',t).
\label{lb28}
\end{equation}
The integral over ${\bf k}$ can be done explicitly (see, e.g., \cite{kindetail}) and we obtain
\begin{equation}
\frac{\partial f}{\partial t}=K_d\frac{\partial}{\partial v_i}\int d{\bf v}' \frac{w^2\delta_{ij}-w_iw_j}{w^3}\left (\frac{\partial}{\partial {v}_{j}}-\frac{\partial}{\partial {v'}_{j}}\right )f({\bf v},t)f({\bf v}',t),
\label{lb29}
\end{equation}
where ${\bf w}={\bf v}-{\bf v}'$ is the relative velocity and $K_d$ is a constant with value $K_3=8\pi^5m\int_0^{+\infty}k^3\hat{u}(k)^2\, dk$ in $d=3$ and  $K_2=8\pi^3 m\int_0^{+\infty}k^2\hat{u}(k)^2\, dk$ in $d=2$. This kinetic equation, that neglects collective effects, is the Landau equation \cite{landau}. We note that the structure of this kinetic equation does not depend on the specific form of potential of interaction. The potential of interaction only appears in factor of the integral in a constant $K_d$ that determines the relaxation time. For a 3D plasma, using $(2\pi)^3\hat{u}(k)=4\pi e^2/m^2k^2$, we get $K_3=(2\pi e^4/m^3)\ln\Lambda$ where $\ln\Lambda=\int_0^{+\infty} dk/k$ is a Coulomb logarithm that has to be regularized with appropriate cut-offs. The large-scale cut-off is the Debye length $\lambda_D^2\sim k_B T/ne^2$ and the small-scale cut-off is the Landau length $\lambda_{L}\sim e^2/m v_m^2\sim e^2/k_B T\sim 1/n\lambda_D^2$ (see Paper II for more details). Therefore $\Lambda\sim \lambda_D/\lambda_{L}\sim n\lambda_D^3\gg 1$ represents the number of electrons in the Debye sphere. For a 2D plasma, using $(2\pi)^2\hat{u}(k)=2\pi e^2/m^2k^2$ and introducing a large-scale cut-off at the Debye length, we obtain $K_2=2\pi e^4/m^3 k_D$. The validity of these heuristic procedures will be further discussed in Paper II. Returning to Eqs. (\ref{lb27}) and (\ref{lb28}), we note that collective effects can be taken into account simply by replacing the bare potential $\hat{u}(k)$ in the Landau equation by a ``dressed'' potential $\hat{u}_{dressed}(k)=\hat{u}(k)/|\epsilon({\bf k},{\bf k}\cdot {\bf v})|$, including the dielectric function, without changing the overall structure of the kinetic equation. Physically, this means that the particles are ``dressed'' by their polarization cloud. In plasma physics, collective effects are important because they account for screening effects and regularize, at the scale of the Debye length, the logarithmic divergence that occurs in the Landau equation. This avoids the introduction of {\it ad hoc} cut-offs.

Systems with long-range interactions are generically spatially inhomogeneous. If we neglect collective effects, we can derive a generalized Landau equation \cite{kandrup1,hb3,kindetail}:
\begin{eqnarray}
\frac{\partial f}{\partial t}+{\bf v}\cdot \frac{\partial f}{\partial {\bf r}}-\frac{N-1}{N} \nabla\Phi\cdot \frac{\partial f}{\partial {\bf v}}=\frac{\partial}{\partial {v}^{\mu}}\int_0^{t} d\tau \int d{\bf
r}_{1}d{\bf v}_1 {F}^{\mu}(1\rightarrow 0){\cal G}(t,t-\tau)\nonumber\\
\times  \left \lbrack {{\cal F}}^{\nu}(1\rightarrow 0)
{\partial\over\partial {v}^{\nu}}+{{\cal F}}^{\nu}(0\rightarrow
1) {\partial\over\partial {v}_{1}^{\nu}}\right
\rbrack f({\bf
r},{\bf v},t-\tau)\frac{f}{m}({\bf r}_1,{\bf v}_1,t-\tau), \label{lb30}
\end{eqnarray}
that is valid for systems that are not necessarily spatially homogeneous, and not necessarily Markovian \footnote{The Markovian approximation may not be justified in every situation. For example, in stellar dynamics, the force auto-correlation function decreases algebraically like $1/t$ \cite{chandrat}, instead of exponentially. Therefore, memory effects are sometimes believed to be important \cite{ps,kandrup1}. In fact, this ``pathology'' is due to the illicit assumption of an infinite homogeneous system. The decay of the correlation functions is more rapid when the spatial inhomogeneity and the finite extent of the system are properly accounted for. In that case, the Markovian approximation is justified \cite{sh}. For the HMF model, the Markovian approximation may be incorrect close to the critical point where the (exponential) decay rate of the force auto-correlation diverges \cite{hb2}.}. For spatially homogeneous Markovian systems,  we recover  the Landau equation (\ref{lb28}) after simple calculations \cite{kindetail}. Equation (\ref{lb30}) shows that the Landau equation has the structure of a Fokker-Planck equation and that the coefficients of diffusion and drift are given by generalized Kubo formulae. If we assume that the system is in a quasi stationary  state (QSS) of the Vlasov equation slowly evolving under the effect of ``collisions'', we can derive a spatially inhomogeneous  Landau  or Lenard-Balescu equation written in angle-action variables \cite{angleaction,kindetail,heyvaerts,newangleaction}. In certain cases, e.g. for 3D self-gravitating systems, we can make a {\it local approximation}. In that case, the kinetic equation (\ref{lb27}) is replaced by
\begin{equation}
\frac{\partial f}{\partial t}+{\bf v}\cdot\frac{\partial f}{\partial {\bf r}}-\nabla\Phi\cdot\frac{\partial f}{\partial {\bf v}}=\pi (2\pi)^{d}m\frac{\partial}{\partial v_i}  \int d{\bf k} \, d{\bf v}'  \, k_ik_j  \frac{\hat{u}(k)^2}{|\epsilon({\bf k},{\bf k}\cdot {\bf v})|^2}\delta\lbrack {\bf k}\cdot ({\bf v}-{\bf v}')\rbrack\left (\frac{\partial}{\partial {v}_{j}}-\frac{\partial}{\partial {v'}_{j}}\right )f({\bf r},{\bf v},t)f({\bf r},{\bf v}',t).
\label{lb31}
\end{equation}
In this equation, the effects of spatial inhomogeneity are kept only in the advection (Vlasov) term, while the collision term is calculated as if the system were spatially homogeneous. If we neglect collective effects, taking $|\epsilon({\bf k},{\bf k}\cdot {\bf v})|=1$, this equation can be rewritten
\begin{equation}
\frac{\partial f}{\partial t}+{\bf v}\cdot\frac{\partial f}{\partial {\bf r}}-\nabla\Phi\cdot\frac{\partial f}{\partial {\bf v}}=K_d\frac{\partial}{\partial v_i}\int d{\bf v}' \frac{w^2\delta_{ij}-w_iw_j}{w^3}\left (\frac{\partial}{\partial {v}_{j}}-\frac{\partial}{\partial {v'}_{j}}\right )f({\bf r},{\bf v},t)f({\bf r},{\bf v}',t).
\label{lb32}
\end{equation}
For a 3D self-gravitating system, using $(2\pi)^3\hat{u}(k)=-4\pi G/k^2$, we get $K_3=2\pi mG^2\ln\Lambda$ where $\ln\Lambda=\int_0^{+\infty} dk/k$ is the Coulomb factor that has to be regularized with appropriate cut-offs. The large-scale cut-off is the Jeans length $\lambda_J^2\sim k_B T/G m^2 n$ (which is of the order of the system size $R$) and the small-scale cut-off is the gravitational Landau length $\lambda_{L}\sim Gm/v_m^2\sim G m^2/k_B T\sim 1/n\lambda_J^2$. Therefore $\Lambda\sim \lambda_J/\lambda_{L}\sim n\lambda_J^3\sim N \gg 1$ represents the number of stars  in the cluster. The Vlasov-Landau kinetic equation (\ref{lb32}) is the fundamental kinetic equation of stellar dynamics \cite{spitzerbook}. For a 2D stellar system, using $(2\pi)^2\hat{u}(k)=-2\pi G/k^2$ and introducing a cut-off at the Jeans length, we obtain $K_2=2\pi G^2 m/ k_J$, but the local approximation is not justified in that case (see Paper II).

\subsection{The relaxation time of the system as a whole}
\label{sec_relaxwhole}

The Lenard-Balescu  equation (\ref{lb27}) is valid at the order $1/N$ so it describes the ``collisional'' evolution of the system on a timescale $\sim N t_D$.  This kinetic equation conserves the mass $M$ and the energy $E$ (which reduces to the kinetic energy for a spatially homogeneous system). It also monotonically increases the Boltzmann entropy $S$ ($H$-theorem) \cite{klimontovich}. The collisional evolution of the system is due to a condition of resonance. This condition of resonance, encapsulated in the $\delta$-function, corresponds to ${\bf k}\cdot {\bf v}'={\bf k}\cdot {\bf v}$ with ${\bf v}'\neq {\bf v}$.

When $d>1$, the condition of resonance can always be satisfied and the Boltzmann distribution (\ref{pvm14}) is the unique steady state of the Lenard-Balescu equation. Therefore, under the effect of ``collisions'', the system reaches the Boltzmann distribution on a relaxation time
 \begin{equation}
t_{R}\sim Nt_D,  \qquad (d>1 \,\, {\rm homogeneous}).
\label{rw1}
\end{equation}
For 3D plasmas, we get from Eq. (\ref{lb29}) an estimate of the relaxation time as $t_R\sim m^2v_m^3/\lbrack ne^4\ln(m v_m^2/e^2 n^{1/3})\rbrack \sim \Lambda^{1/2}\lambda_D^{3/2}m^{1/2}/e\ln\Lambda$, leading to $t_R\sim(\Lambda/\ln\Lambda)t_D$. For 2D plasmas, we obtain $t_R\sim m^2v_m^3k_D/ne^4\sim m^{3/2}v_m^2/n^{1/2}e^3\sim \Lambda^{1/2}\lambda_D m^{1/2}/e$, leading to $t_R\sim \Lambda t_D$. In both cases, $\Lambda=n\lambda_D^d$ represents the number of electrons in a Debye sphere and $t_D\sim\omega_P^{-1}$, where $\omega_P\sim (ne^2/m)^{1/2}\sim v_m/\lambda_D$ is the plasma pulsation, is a measure of the dynamical time. For 3D stellar systems, we get from Eq. (\ref{lb32}) an estimate of the relaxation time as $t_R\sim v_m^3/\lbrack nm^2G^2\ln (v_m^2/G m n^{1/3})\rbrack \sim N^{1/2}\lambda_J^{3/2}/m^{1/2}G^{1/2}\ln N$ leading to $t_R\sim (N/\ln N)t_D$.  For 2D stellar systems, we obtain $t_R\sim v_m^3k_J/n m^2 G^2\sim v_m^2/n^{1/2}m^{3/2}G^{3/2}\sim N^{1/2}\lambda_J/m^{1/2}G^{1/2}\sim N t_D$. In both cases, $\Lambda=n\lambda_J^d\sim N$ represents the number of stars in the cluster  and $t_D\sim\omega_G^{-1}$, where $\omega_G\sim (G\rho)^{1/2}\sim v_m/\lambda_J$ is the gravitational pulsation, is a measure of the dynamical time. We have just given the scaling of the relaxation time as the prefactor depends on the precise way the relaxation time is defined.

For spatially homogeneous one dimensional systems, there is no resonance and the Lenard-Balescu collision term vanishes. Indeed, the kinetic equation  (\ref{lb27}) reduces to
 \begin{equation}
\frac{\partial f}{\partial t}=2\pi^2 m\frac{\partial}{\partial v}  \int d{k} \, d{v}'  \, |k|  \frac{\hat{u}(k)^2}{|\epsilon({k},k v)|^2}\delta ({v}-{v}')\left (\frac{\partial}{\partial {v}}-\frac{\partial}{\partial {v'}}\right )f({v},t)f({v}',t)=0,
\label{rw2}
\end{equation}
so the distribution function does not evolve at all on a timescale $\sim N t_D$. Therefore, the kinetic theory predicts no thermalization to a Maxwellian at first order in $1/N$. The maxwellization is at least a second order effect in $1/N$, and consequently a very slow process. The relaxation time satisfies
 \begin{equation}
t_{R}>Nt_D  \qquad (d=1  \,\, {\rm homogeneous}).
\label{rw3}
\end{equation}
This result has been known for a long time in plasma physics \cite{feix,kp} and was rediscovered recently in the context of the HMF model \cite{bd,cvb}.  Since the relaxation process is due to more complex correlations than simply two-body collisions, we have to develop the kinetic theory at higher orders (taking  three-body, four-body,... correlation functions into account) in order to obtain the relaxation time. If the collision term does not vanish at the next order of the expansion in powers of $1/N$, the kinetic theory would imply a relaxation time of the order $N^2t_D$. This quadratic scaling, conjectured in \cite{feix}, is observed for 1D plasmas  that are spatially homogeneous \cite{dawson,rouetfeix}. In that case, the relaxation is caused by three-body correlations. However, the problem could be more complicated and yield a larger relaxation time like $e^N t_D$, as observed in \cite{campa} for the permanently homogeneous HMF model (but other authors \cite{private} find the more natural $N^2$ scaling). In fact, it is not even granted that the system will ever relax towards statistical equilibrium; the evolution may be non-ergodic (the mixing by ``collisions'' may be inefficient).

For spatially inhomogeneous systems, since there are potentially more resonances (as can be seen from the Lenard-Balescu equation written in angle-action variables \cite{angleaction,kindetail,heyvaerts,newangleaction}), the relaxation time could be reduced and approach the natural scaling
\begin{equation}
t_{R}\sim Nt_D\qquad ({\rm inhomogeneous}),
\label{rw4}
\end{equation}
predicted by the first order kinetic theory \cite{angleaction,kindetail}. This linear scaling is observed for 1D self-gravitating systems that are spatially inhomogeneous \cite{brucemiller,gouda,valageas2,joyce} and for the permanently inhomogeneous HMF model \cite{ruffoN}. However, very little is known concerning the properties of the spatially inhomogeneous Lenard-Balescu equation and its convergence (or not) towards the Boltzmann distribution. It could approach the Boltzmann distribution (since entropy increases) without reaching it exactly if, at some point, the condition of resonance cannot be satisfied anymore.

Finally, if we consider a 1D system that is initially spatially {homogeneous} but, slowly evolving under the effect of collisions, becomes Vlasov unstable and undergoes a dynamical phase transition  \cite{campachav} making it spatially inhomogeneous, we expect a relaxation time scaling like $N^{\delta}$ with $1<\delta<2$, intermediate between the scalings (\ref{rw3}) and (\ref{rw4}) of permanently homogeneous and permanently inhomogeneous systems. Such a scaling, with $\delta=1.7$, has been observed in \cite{yamaguchi} for the HMF model experiencing a dynamical phase transition.

The same type of results are found for 2D point vortices \cite{collective}.

\section{Stochastic process of a test particle: Diffusion and friction}
\label{sec_stoch}

\subsection{The Fokker-Planck equation}
\label{sec_fp}

In the previous section, we have studied the evolution of the system as a whole. We now consider the relaxation of a test particle in a bath of field particles with a steady distribution function $f({\bf v})$. We assume that the field particles are either (i) at statistical equilibrium with the Boltzmann distribution (thermal bath), in which case their distribution does not change at all, or (ii) in the case of one-dimensional systems, in any stable steady state of the Vlasov equation (as we have just seen, in $d=1$, this profile does not change on a timescale of order $Nt_D$). We assume that the test particle has an initial velocity ${\bf v}_0$ and we study how it progressively acquires the distribution of the bath due to ``collisions'' with the field particles. As we shall see, the test particle has a stochastic motion and the evolution of the distribution function $P({\bf v},t)$, the probability density that the test particle has a velocity ${\bf v}$ at time $t$,  is governed by a Fokker-Planck equation involving a diffusion term and a friction term that can be analytically obtained. The Fokker-Planck equation may then be solved with the initial condition $P({\bf v},t_0)=\delta({\bf v}-{\bf v}_0)$ to yield $P({\bf v},t)$. In plasma physics, this test particle approach has been developed by Hubbard  \cite{hubbard1} who properly took collective effects into account. The derivation that we give below is slightly different and follows \cite{cdr} closely.

The equations of motion of the test particle are
\begin{equation}
\label{fp1}
\frac{d{\bf r}}{dt}={\bf v},\qquad \frac{d{\bf v}}{dt}=-\nabla \delta\Phi({\bf r},t).
\end{equation}
When $N\rightarrow +\infty$, the fluctuations of the potential can be neglected and the test particle follows a linear trajectory at constant velocity ${\bf v}$. At the order $1/N$, the fluctuations $\delta\Phi$ of the potential must be taken into account and the particle has a stochastic motion. Equations (\ref{fp1}) can be formally integrated into
\begin{equation}
\label{fp2}
{\bf r}(t)={\bf r}+\int_{0}^{t} {\bf v}(t')\, dt',
\end{equation}
\begin{equation}
\label{fp3}
{\bf v}(t)={\bf v}-\int_{0}^{t}\nabla\delta\Phi({\bf r}(t'),t')\, dt',
\end{equation}
where we have assumed that, initially, the test particle is at $({\bf r},{\bf v})$. Note that the ``initial'' time considered here does not necessarily coincide with the original time $t_0$ mentioned above. Since the fluctuation $\delta\Phi$ of the potential is a small quantity, the foregoing equations can be solved perturbatively. At the order $1/N$, which corresponds to quadratic order in $\delta\Phi$, we get for the velocity
\begin{eqnarray}
\label{fp4}
v_{i}(t)=v_{i}-\int_{0}^{t}dt'\, \frac{\partial\delta\Phi}{\partial x_{i}}({\bf r}+{\bf v}t',t')
+\int_{0}^{t}dt'\int_{0}^{t'}dt''\int_{0}^{t''}dt'''\, \frac{\partial^{2}\delta\Phi}{\partial x_{i}\partial x_{j}}({\bf r}+{\bf v}t', t')\frac{\partial\delta\Phi}{\partial x_{j}} ({\bf r}+{\bf v}t''', t''').
\end{eqnarray}
As the changes in the velocity are small, the dynamics of the test particle can be represented by a stochastic process governed by a Fokker-Planck equation \cite{risken}. If we denote by $P({\bf v},t)$ the density probability that the test particle has a velocity ${\bf v}$ at time $t$, the general form of this equation is
\begin{equation}
\label{fp5}
\frac{\partial P}{\partial t}=\frac{\partial^{2}}{\partial v_{i}\partial v_{j}}\left (D_{ij}P\right )-\frac{\partial}{\partial v_{i}}\left (PA_{i}\right ).
\end{equation}
The diffusion tensor and the friction force are defined by
\begin{equation}
\label{fp6}
D_{ij}({\bf v})=\lim_{t\rightarrow +\infty}\frac{1}{2t} \langle (v_{i}(t)-v_{i})(v_{j}(t)-v_{j})\rangle,
\end{equation}
\begin{equation}
\label{fp7}
A_{i}({\bf v})=\lim_{t\rightarrow +\infty}\frac{1}{t} \langle v_{i}(t)-v_{i}\rangle.
\end{equation}
In writing these limits, we have implicitly assumed that the time $t$ is long with respect to the fluctuation time but short with respect to the relaxation time (of order $Nt_D$), so the expression (\ref{fp4}) can be used to evaluate Eqs. (\ref{fp6}) and (\ref{fp7}). As shown in our previous papers \cite{hb2,hb4,kindetail}, it is relevant to rewrite the Fokker-Planck equation in the alternative form
\begin{equation}
\label{fp8}
\frac{\partial P}{\partial t}=\frac{\partial}{\partial v_{i}} \left (D_{ij}\frac{\partial P}{\partial
v_{j}}- P F_i^{pol}\right ).
\end{equation}
The total friction is
\begin{equation}
\label{fp9}
A_{i}=F_{i}^{pol}+\frac{\partial D_{ij}}{\partial v_{j}},
\end{equation}
where ${\bf F}_{pol}$ is the friction due to the  polarization \footnote{As shown in \cite{kalnajs,kandrup2,hb4}, the friction by  polarization ${\bf F}_{pol}$ arises from the retroaction of the field particles to the perturbation caused by the test particle. It represents, however, only one component of the  dynamical friction ${\bf A}$, the other component being $\partial_j D_{ij}$. When collective effects are neglected, these two components are equal \cite{hb4}. This observation  explains the difference, by a factor of $2$ (for equal mass particles), between the calculations of Kalnajs \cite{kalnajs} and Kandrup \cite{kandrup2}, giving ${\bf F}_{pol}$,  and the calculation of Chandrasekhar \cite{chandra1}, giving ${\bf A}$. On the other hand, the diffusion tensor $D_{ij}$ is due to the fluctuations of the force. It can be obtained from the Kubo formula \cite{kandrup2,hb4}. For a thermal bath, it is connected to the friction by polarization by the Einstein relation (see Sec. \ref{sec_einstein}).}, while the second term is due to the variations of the diffusion tensor with the velocity ${\bf v}$. As we shall see, this decomposition arises naturally in the following analysis. The two expressions (\ref{fp5}) and (\ref{fp8}) have their own interest. The expression (\ref{fp5}) where the diffusion tensor is placed after the second derivative $\partial^2(DP)$ involves the total (dynamical) friction ${\bf A}$ and the expression (\ref{fp8}) where the diffusion tensor is placed between the derivatives $\partial D\partial P$ isolates the friction by polarization ${\bf F}_{pol}$.  We shall see in Sec. \ref{sec_conn} that this second form is directly related to the Lenard-Balescu equation (\ref{lb27}). It has therefore a clear physical interpretation. We shall now calculate the diffusion tensor and the friction term from Eqs. (\ref{fp6}) and (\ref{fp7}), using the results of Sec. \ref{sec_lb} that allow us to take collective effects into account.

\subsection{The diffusion tensor}
\label{sec_diff}

We first compute the diffusion tensor defined by Eq. (\ref{fp6}). Using Eq. (\ref{fp4}), we see that it is given, at the order $1/N$, by
\begin{equation}
D_{ij}=\frac{1}{2t}\int_{0}^{t}dt'\int_{0}^{t}dt'' \, \left\langle \frac{\partial\delta\Phi}{\partial x_i}({\bf r}+{\bf v}t', t')\frac{\partial\delta\Phi}{\partial x_j}({\bf r}+{\bf v}t'', t'')\right\rangle.
\label{diff1}
\end{equation}
By the inverse Fourier-Laplace transform, we have
\begin{eqnarray}
\left\langle \frac{\partial\delta\Phi}{\partial x_i}({\bf r}+{\bf v}t', t')\frac{\partial\delta\Phi}{\partial x_j}({\bf r}+{\bf v}t'', t'')\right\rangle=-\int d{\bf k}\int_{\cal C}\frac{d\omega}{2\pi}\int d{{\bf k}'}\int_{\cal C}\frac{d\omega'}{2\pi} \, k_i k'_j e^{i{\bf k}\cdot ({\bf r}+{\bf v} t')}e^{-i\omega t'}e^{i {\bf k}'\cdot ({\bf r}+{\bf v} t'')}e^{-i\omega' t''} \nonumber\\
\times\langle \delta\tilde\Phi({\bf k},\omega)\delta\tilde\Phi({\bf k}',\omega')\rangle.\qquad
\label{diff2}
\end{eqnarray}
Substituting Eq. (\ref{lb18}) in Eq. (\ref{diff2}), and carrying out the integrals over ${\bf k}'$, $\omega'$ and $\omega$, we end up with the result
\begin{eqnarray}
\left\langle \frac{\partial\delta\Phi}{\partial x_i}({\bf r}+{\bf v}t', t')\frac{\partial\delta\Phi}{\partial x_j}({\bf r}+{\bf v}t'', t'')\right\rangle=(2\pi)^d m \int d{\bf k}\, d{\bf v}' \, k_i k_j e^{i{\bf k}\cdot ({\bf v}-{\bf v}')(t'-t'')}\frac{\hat{u}(k)^2}{|\epsilon({\bf k},{\bf k}\cdot {\bf v}')|^2}f({\bf v}').
\label{diff3}
\end{eqnarray}
This expression shows that the correlation function appearing in Eq. (\ref{diff1}) is an even function of $t'-t''$. Using the identity
\begin{eqnarray}
\int_{0}^{t}dt'\int_{0}^{t}dt''\, f(t'-t'')=2\int_{0}^{t}dt'\int_{0}^{t'}dt''\, f(t'-t'')=2\int_{0}^{t}ds\, (t-s)f(s),
\label{diff4}
\end{eqnarray}
we find, for $t\rightarrow +\infty$, that
\begin{equation}
D_{ij}=\int_{0}^{+\infty} \left\langle \frac{\partial\delta\Phi}{\partial x_i}({\bf r},0)\frac{\partial\delta\Phi}{\partial x_j}({\bf r}+{\bf v}s, s)\right\rangle\, ds.
\label{diff5}
\end{equation}
This is the Kubo formula for our problem. Replacing the correlation function by its expression (\ref{diff3}), we get
\begin{equation}
D_{ij}=(2\pi)^d m \int_{0}^{+\infty}ds\int d{\bf k}\, d{\bf v}' \, k_i k_j e^{i{\bf k}\cdot ({\bf v}-{\bf v}')s}\frac{\hat{u}(k)^2}{|\epsilon({\bf k},{\bf k}\cdot {\bf v}')|^2}f({\bf v}').
\label{diff6}
\end{equation}
Making the change of variables $s\rightarrow -s$ and ${\bf k}\rightarrow -{\bf k}$, we see that we can replace $\int_{0}^{+\infty}ds$ by $(1/2)\int_{-\infty}^{+\infty}ds$ in Eq. (\ref{diff6}). Then, using the identity
\begin{equation}
\delta(\omega)=\int_{-\infty}^{+\infty} e^{i \omega t}\, \frac{dt}{2\pi},
\label{delta}
\end{equation}
we obtain the final expression
\begin{equation}
D_{ij}=\pi(2\pi)^d m \int d{\bf k}\, d{\bf v}' \, k_i k_j \frac{\hat{u}(k)^2}{|\epsilon({\bf k},{\bf k}\cdot {\bf v})|^2}\delta\lbrack {\bf k}\cdot ({\bf v}-{\bf v}')\rbrack f({\bf v}').
\label{diff7}
\end{equation}

\subsection{The friction due to the polarization}
\label{sec_ess}

We now compute the friction force defined by Eq. (\ref{fp7}). We need to keep terms up to order $1/N$. From Eq. (\ref{fp4}), the first term to compute is
\begin{equation}
{\bf A}_{I}=-\frac{1}{t}\int_{0}^{t}dt'\, \left\langle \nabla\delta\Phi({\bf r}+{\bf v} t',t')\right\rangle.
\label{ess1}
\end{equation}
By the inverse Fourier-Laplace transform, we have
\begin{equation}
\left\langle\nabla\delta\Phi({\bf r}+{\bf v} t',t')\right\rangle=i\int d{\bf k}\int_{\cal C}\frac{d\omega}{2\pi} \, {\bf k}  e^{i {\bf k}\cdot ({\bf r}+{\bf v}t')}e^{-i\omega t'}\langle \delta\tilde\Phi({\bf k},\omega)\rangle.
\label{ess2}
\end{equation}
Using Eq. (\ref{lb10}), we find that
\begin{equation}
\langle \delta\tilde\Phi({\bf k},\omega)\rangle=(2\pi)^d \frac{\hat{u}(k)}{\epsilon({\bf k},\omega)}  \int d{\bf v}'\, \frac{\langle \delta\hat f({\bf k},{\bf v}',0)\rangle}{i({\bf k}\cdot {\bf v}'-\omega)}.
\label{ess3}
\end{equation}
Now, using the fact that the test particle is initially located in $({\bf r},{\bf v})$, so that $\langle \delta f({\bf r}',{\bf v}',0)\rangle=m\delta({\bf r}'-{\bf r})\delta({\bf v}'-{\bf v})$,  we obtain from Eq. (\ref{lb8}) the result
\begin{equation}
\langle \delta\hat f({\bf k},{\bf v}',0)\rangle=\frac{1}{(2\pi)^d}m\, e^{-i{\bf k}\cdot {\bf r}}\delta({\bf v}'-{\bf v}).
\label{ess4}
\end{equation}
Substituting these expressions in Eq. (\ref{ess2}), we get
\begin{equation}
\left\langle\nabla\delta\Phi({\bf r}+{\bf v} t',t')\right\rangle={m}\int d{\bf k}\int_{\cal C}\frac{d\omega}{2\pi} \, {\bf k}  e^{i ({\bf k}\cdot {\bf v}-\omega) t'}\frac{\hat{u}(k)}{\epsilon({\bf k},\omega)} \frac{1}{{\bf k}\cdot {\bf v}-\omega}.
\label{ess5}
\end{equation}
Therefore, the friction term (\ref{ess1}) is given by
\begin{equation}
{\bf A}_{I}=-m\frac{1}{t}\int_{0}^{t}dt'\int d{\bf k}\int_{\cal C}\frac{d\omega}{2\pi} \, {\bf k}  e^{i ({\bf k}\cdot {\bf v}-\omega) t'}\frac{\hat{u}(k)}{\epsilon({\bf k},\omega)}  \frac{1}{{\bf k}\cdot {\bf v}-\omega}.
\label{ess6}
\end{equation}
We now use the Landau prescription $\omega\rightarrow \omega+i0^{+}$ and the Plemelj formula (\ref{lb21}) to evaluate the integral over $\omega$.   The term corresponding to the imaginary part in the  Plemelj formula is
\begin{equation}
{\bf A}_{I}^{(a)}=-i m \pi \frac{1}{t}\int_{0}^{t}dt'\int d{\bf k}\int\frac{d\omega}{2\pi} \, {\bf k}  e^{i ({\bf k}\cdot {\bf v}-\omega) t'}\frac{\hat{u}(k)}{\epsilon({\bf k},\omega)}  \delta({\bf k}\cdot {\bf v}-\omega).
\label{ess7}
\end{equation}
Integrating over $\omega$ and $t'$, we obtain
\begin{equation}
{\bf A}_{I}^{(a)}=-\frac{m}{2}\int d{\bf k} \, {\bf k}  \, \frac{\hat{u}(k)}{|\epsilon({\bf k},{\bf k}\cdot {\bf v})|^2} {\rm Im} \left\lbrack \epsilon({\bf k},{\bf k}\cdot {\bf v})\right\rbrack.
\label{ess8}
\end{equation}
The term corresponding to the real part in the Plemelj formula  is
\begin{equation}
{\bf A}_{I}^{(b)}=-m\frac{1}{t}\int_{0}^{t}dt'\int d{\bf k}\, {\cal P}\int_{-\infty}^{+\infty}\frac{d\omega}{2\pi} \, {\bf k}  e^{i ({\bf k}\cdot {\bf v}-\omega) t'}\frac{\hat{u}(k)}{\epsilon({\bf k},\omega)} \frac{1}{{\bf k}\cdot {\bf v}-\omega}.
\label{ess10}
\end{equation}
Integrating over $t'$, we can convert this expression to the form
\begin{equation}
{\bf A}_{I}^{(b)}=i m\int d{\bf k}\, {\cal P}\int_{-\infty}^{+\infty}\frac{d\omega}{2\pi} \, {\bf k}\frac{\hat{u}(k)}{\epsilon({\bf k},\omega)} \frac{1}{({\bf k}\cdot {\bf v}-\omega)^2}\frac{1}{t} \left\lbrace i\sin \left\lbrack ({\bf k}\cdot {\bf v}-\omega)t\right\rbrack+\cos \left\lbrack ({\bf k}\cdot {\bf v}-\omega)t\right\rbrack-1\right\rbrace.
\label{ess11}
\end{equation}
For $t\rightarrow +\infty$, using the identity
\begin{equation}
\lim_{t\rightarrow +\infty}\frac{1-\cos(tx)}{t x^2}=\pi\delta(x),
\label{ess12}
\end{equation}
and integrating over $\omega$, we find that ${\bf A}_{I}^{(b)}$ is given by Eq. (\ref{ess8}), just like ${\bf A}_{I}^{(a)}$. Therefore, writing ${\bf A}_I={\bf A}_{I}^{(a)}+{\bf A}_{I}^{(b)}=2{\bf A}_{I}^{(a)}$, we obtain
\begin{equation}
{\bf A}_{I}=-m\int d{\bf k} \, {\bf k}  \, \frac{\hat{u}(k)}{|\epsilon({\bf k},{\bf k}\cdot {\bf v})|^2} {\rm Im} \left\lbrack \epsilon({\bf k},{\bf k}\cdot {\bf v})\right\rbrack.
\label{ess8b}
\end{equation}
Finally, using Eq. (\ref{lb25}), we find that
\begin{equation}
{A}_i^{I}=\pi (2\pi)^d m\int d{\bf k}\, d{\bf v}'\, k_i k_j \frac{\hat{u}(k)^2}{|\epsilon({\bf k},{\bf k}\cdot {\bf v})|^2}\delta\lbrack {\bf k}\cdot ({\bf v}- {\bf v}')\rbrack \frac{\partial f}{\partial {v_j}'}({\bf v}').
\label{ess14}
\end{equation}
As we shall see, ${\bf A}_I$  corresponds to the friction due to the polarization denoted  ${\bf F}_{pol}$ in Eq. (\ref{fp9}).

{\it Remark:} We can obtain Eq. (\ref{ess14}) in a more direct manner from Eq. (\ref{ess5}) by using the contour of integration shown in Fig. 9 of \cite{pitaevskii}. In that case, the integral over $\omega$ is just $-2\pi i$ times the sum of the residues at the poles of the integrand in Eq. (\ref{ess5}). The poles corresponding to the zeros of the dielectric function $\epsilon({\bf k},\omega)$ give a contribution that rapidly decays with time since ${\rm Im}(\omega)<0$ (the system is Vlasov stable). Keeping only the contribution of the pole $\omega={\bf k}\cdot {\bf v}$ that does not decay in time we obtain
\begin{equation}
\left\langle\nabla\delta\Phi({\bf r}+{\bf v} t',t')\right\rangle=i {m}\int d{\bf k} \, {\bf k} \frac{\hat{u}(k)}{\epsilon({\bf k},{\bf k}\cdot {\bf v})}.
\label{ess5bis}
\end{equation}
Substituting this result in Eq. (\ref{ess1}), we get Eq. (\ref{ess8b}) then Eq. (\ref{ess14}).

\subsection{The part of the friction due to the inhomogeneity of the diffusion coefficient}
\label{sec_add}

In the evaluation of the total friction, at the order $1/N$, the second term to compute is
\begin{equation}
A_{i}^{II}=\frac{1}{t} \int_{0}^{t}dt'\int_{0}^{t'}dt''\int_{0}^{t''}dt'''\, \left\langle\frac{\partial^{2}\delta\Phi}{\partial x_{i}\partial x_{j}}({\bf r}+{\bf v}t', t')\frac{\partial\delta\Phi}{\partial x_{j}} ({\bf r}+{\bf v}t''', t''')\right\rangle.
\label{add1}
\end{equation}
By the inverse Fourier-Laplace transform, we have
\begin{eqnarray}
\left\langle\frac{\partial^{2}\delta\Phi}{\partial x_{i}\partial x_{j}}({\bf r}+{\bf v}t', t')\frac{\partial\delta\Phi}{\partial x_{j}} ({\bf r}+{\bf v}t''',  t''')\right\rangle=-i\int d{\bf k}\int_{\cal C}\frac{d\omega}{2\pi}\int d{\bf k}'\int_{\cal C}\frac{d\omega'}{2\pi} \, k_{i}k_{j}k'_{j}   e^{i {\bf k}\cdot ({\bf r}+{\bf v} t')}e^{-i\omega t'}\nonumber\\
\times e^{i {\bf k'}({\bf r}+{\bf v} t''')}e^{-i\omega' t'''} \langle \delta\tilde\Phi({\bf k},\omega)\delta\tilde\Phi({\bf k}',\omega')\rangle.
\label{add2}
\end{eqnarray}
Substituting Eq. (\ref{lb18}) in Eq. (\ref{add2}), and carrying out the integrals over ${\bf k}'$, $\omega'$ and $\omega$, we end up with the result
\begin{eqnarray}
\left\langle\frac{\partial^{2}\delta\Phi}{\partial x_{i}\partial x_{j}}({\bf r}+{\bf v}t', t')\frac{\partial\delta\Phi}{\partial x_{j}} ({\bf r}+{\bf v}t''',  t''')\right\rangle=i\, (2\pi)^d m \int d{\bf k}\,  d{\bf v}' k_i k^2  e^{i {\bf k}\cdot ({\bf v}-{\bf v}')(t'-t''')} \frac{\hat{u}(k)^2}{|\epsilon({\bf k},{\bf k}\cdot{\bf v}')|^2}f({\bf v}').
\label{add3}
\end{eqnarray}
This expression shows that the correlation function appearing in Eq. (\ref{add1}) is an odd function of $t'-t'''$. Using the identity
\begin{eqnarray}
\int_{0}^{t'}dt''\int_{0}^{t''}dt'''\, f(t'-t''')=\int_{0}^{t'}dt''' \, (t'-t''')f(t'-t'''),
\label{add4}
\end{eqnarray}
we find that
\begin{equation}
A_{i}^{II}=i \, (2\pi)^d m\frac{1}{t}\int_{0}^{t}dt'\int_{0}^{t'}dt'''\int d{\bf k}\, d{\bf v}' \,  k_i k^2 (t'-t''') e^{i {\bf k}\cdot ({\bf v}-{\bf v}')(t'-t''')}\frac{\hat{u}(k)^2}{|\epsilon({\bf k},{\bf k}\cdot{\bf v}')|^2} f({\bf v}').
\label{add5}
\end{equation}
This can be rewritten
\begin{eqnarray}
A_{i}^{II}=(2\pi)^d m\frac{1}{t}\frac{\partial}{\partial v_{j}}\int_{0}^{t}dt'\int_{0}^{t'}dt'''\int d{\bf k}\,  d{\bf v}' \,  k_i k_j  e^{i {\bf k}\cdot ({\bf v}-{\bf v}')(t'-t''')}\frac{\hat{u}(k)^2}{|\epsilon({\bf k},{\bf k}\cdot{\bf v}')|^2} f({\bf v}').
\label{add7}
\end{eqnarray}
Since the integrand only depends on $t'-t'''$, using the identity (\ref{diff4}), we obtain for $t\rightarrow +\infty$,
\begin{eqnarray}
A_{i}^{II}=(2\pi)^d m\frac{\partial}{\partial v_{j}}\int_{0}^{+\infty}ds\int d{\bf k}\, d{\bf v}' \, k_i k_j e^{i {\bf k}\cdot ({\bf v}-{\bf v}')s} \frac{\hat{u}(k)^2}{|\epsilon({\bf k},{\bf k}\cdot{\bf v}')|^2} f({\bf v}').
\label{add8}
\end{eqnarray}
Therefore, similarly to the Kubo formula (\ref{diff5}) for the diffusion coefficient $D_{ij}$, the friction ${\bf A}_{II}$ can be written
\begin{equation}
A_{i}^{II}=\int_{0}^{+\infty} \left\langle \frac{\partial\delta\Phi}{\partial x_j}({\bf r},0)\frac{\partial^2\delta\Phi}{\partial x_i\partial x_j}({\bf r}+{\bf v}s, s)\right\rangle s\, ds.
\label{diff5vf}
\end{equation}
Making the change of variables $s\rightarrow -s$ and ${\bf k}\rightarrow -{\bf k}$, we see that we can replace $\int_{0}^{+\infty}ds$ by $(1/2)\int_{-\infty}^{+\infty}ds$ in Eq. (\ref{add8}). Then, using the identity (\ref{delta}), we obtain the expression
\begin{eqnarray}
A_{i}^{II}=\pi (2\pi)^d m\frac{\partial}{\partial v_{j}}\int d{\bf k}\,  d{\bf v}' \,  k_i k_j \frac{\hat{u}(k)^2}{|\epsilon({\bf k},{\bf k}\cdot{\bf v})|^2} \delta\lbrack {\bf k}\cdot ({\bf v}-{\bf v}')\rbrack f({\bf v}').
\label{add9a}
\end{eqnarray}
This expression involves the diffusion tensor (\ref{diff7}), so that finally
\begin{eqnarray}
A_{i}^{II}=\frac{\partial D_{ij}}{\partial v_{j}}.
\label{add9}
\end{eqnarray}

{\it Remark:} we can obtain this identity in a more direct manner by taking the partial derivative of Eq. (\ref{diff5}) with respect to $v_j$ (using the fact that $D_{ij}=D_{ji}$) and comparing with Eq. (\ref{diff5vf}).

\subsection{Connection between the Lenard-Balescu equation (\ref{lb27}) and the Fokker-Planck equation (\ref{fp8})}
\label{sec_conn}

We have established that the diffusion tensor and the friction force are given by
\begin{equation}
D_{ij}=\pi(2\pi)^d m \int d{\bf k}\, d{\bf v}' \, k_i k_j \frac{\hat{u}(k)^2}{|\epsilon({\bf k},{\bf k}\cdot {\bf v})|^2}\delta\lbrack {\bf k}\cdot ({\bf v}-{\bf v}')\rbrack f({\bf v}').
\label{diff7again}
\end{equation}
\begin{equation}
A_i=\pi (2\pi)^d m\int d{\bf k}\, d{\bf v}'\, k_i k_j \frac{\hat{u}(k)^2}{|\epsilon({\bf k},{\bf k}\cdot {\bf v})|^2}\delta\lbrack {\bf k}\cdot ({\bf v}- {\bf v}')\rbrack \frac{\partial f}{\partial {v_j}'}({\bf v}')+\frac{\partial D_{ij}}{\partial v_{j}}.
\label{add20again}
\end{equation}
Comparing Eq. (\ref{add20again}) with Eq. (\ref{fp9}), we see that the friction by polarization is
\begin{equation}
{F}_i^{pol}=\pi (2\pi)^d m\int d{\bf k}\, d{\bf v}'\, k_i k_j \frac{\hat{u}(k)^2}{|\epsilon({\bf k},{\bf k}\cdot {\bf v})|^2}\delta\lbrack {\bf k}\cdot ({\bf v}- {\bf v}')\rbrack \frac{\partial f}{\partial {v_j}'}({\bf v}').
\label{ess14again}
\end{equation}
On the other hand, using an integration by parts in the first term of Eq. (\ref{add20again}), the total friction can be written
\begin{equation}
{A}_i=\pi (2\pi)^d m\int d{\bf k}d{\bf v}'\, k_i k_j f({\bf v}') \left (\frac{\partial}{\partial {v}_j}-\frac{\partial}{\partial {v'}_j}\right ) \delta\lbrack {\bf k}\cdot ({\bf v}- {\bf v}')\rbrack \frac{\hat{u}(k)^2}{|\epsilon({\bf k},{\bf k}\cdot {\bf v})|^2}.
\label{conn4}
\end{equation}
The foregoing expressions of the diffusion tensor and friction force, taking collective effects into account, were first derived by Hubbard \cite{hubbard1}. 

Using Eqs. (\ref{diff7again}) and (\ref{ess14again}), we find that the Fokker-Planck equation (\ref{fp8}) becomes
\begin{equation}
\frac{\partial P}{\partial t}=\pi (2\pi)^{d}m\frac{\partial}{\partial v_i}  \int d{\bf k} \, d{\bf v}'  \, k_ik_j   \frac{\hat{u}(k)^2}{|\epsilon({\bf k},{\bf k}\cdot {\bf v})|^2}\delta\lbrack {\bf k}\cdot ({\bf v}-{\bf v}')\rbrack\left (\frac{\partial}{\partial {v}_j}-\frac{\partial}{\partial {v'}_j}\right )P({\bf v},t)f({\bf v}').
\label{conn5}
\end{equation}
If the system is spatially inhomogeneous, but if a local approximation can be justified, the Fokker-Planck equation (\ref{conn5}) is replaced by
\begin{equation}
\frac{\partial P}{\partial t}+{\bf v}\cdot\frac{\partial P}{\partial {\bf r}}-\nabla\Phi({\bf r})\cdot\frac{\partial P}{\partial {\bf v}}=\pi (2\pi)^{d}m\frac{\partial}{\partial v_i}  \int d{\bf k} \, d{\bf v}'  \, k_ik_j  \frac{\hat{u}(k)^2}{|\epsilon({\bf k},{\bf k}\cdot {\bf v})|^2}\delta\lbrack {\bf k}\cdot ({\bf v}-{\bf v}')\rbrack\left (\frac{\partial}{\partial {v}_{j}}-\frac{\partial}{\partial {v'}_{j}}\right )P({\bf r},{\bf v},t)f({\bf r},{\bf v}'),
\label{conn5loc}
\end{equation}
where $\Phi({\bf r})$ is the static potential produced by $f({\bf r},{\bf v})$. The form of Eq. (\ref{conn5}) is very similar to the form of Eq. (\ref{lb27}). This shows that the Fokker-Planck equation (\ref{conn5}), with the diffusion coefficient (\ref{diff7again}) and the friction (\ref{ess14again}), can be directly obtained from the Lenard-Balescu equation (\ref{lb27}) by replacing the time dependent distribution $f({\bf v}',t)$ by the {\it static} distribution $f({\bf v}')$ of the bath. This procedure transforms an integro-differential equation (\ref{lb27}) into a  differential equation (\ref{conn5}).  Although natural, the rigorous justification of this procedure requires the detailed calculation that we have given here. In fact, we can understand this result in the following manner. Equations (\ref{lb27}) and (\ref{conn5}) govern the evolution of the distribution function of a test particle (described by the coordinate ${\bf v}$) interacting with field particles (described by the running coordinate ${\bf v}'$). In Eq. (\ref{lb27}), all the particles are equivalent so the distribution of the field particles $f({\bf v}',t)$ changes with time exactly like the distribution of the test particle $f({\bf v},t)$. In Eq. (\ref{conn5}), the test particle and the field particles are not equivalent since the field particles form a ``bath''. The field particles have a steady (given) distribution $f({\bf v}')$ while the distribution of the test particle $f({\bf v},t)=N m P({\bf v},t)$ changes with time. As a result, the Fokker-Planck equation (\ref{conn5}) derived by Hubbard \cite{hubbard1} can be deduced from the kinetic equation (\ref{lb27}) derived by Lenard \cite{lenard} and Balescu \cite{balescu2} and {\it vice versa}.

\subsection{The relaxation time of a test particle in a bath}
\label{sec_rbath}

The derivation of the Fokker-Planck equation (\ref{conn5}), relying on a {\it bath approximation}, assumes that the distribution of the field particles is ``frozen'' so that  $f({\bf v})$ does not change in time, at least on the timescale $Nt_D$ corresponding to the relaxation time of the test particle (see below). This is always true for a thermal bath, corresponding to a distribution  at statistical equilibrium (Boltzmann), because it does not evolve at all. For $d>1$, due to collisions, any other distribution function relaxes towards the Boltzmann distribution on a timescale $Nt_D$. Therefore, for $d>1$, only the Boltzmann distribution can form a bath (the other distributions change in time). The situation is different in $d=1$ since a Vlasov stable distribution function does not change on a timescale of order $Nt_D$. Therefore, any Vlasov stable distribution function can be considered as ``frozen'' on this timescale and can form a bath.

Recalling that $u\sim 1/N$ and $f\sim N$, the Fokker-Planck operator in Eqs. (\ref{conn5}) scales like $1/N$. Therefore, the distribution $P({\bf v},t)$ of the test particle relaxes towards the distribution $f({\bf v})$ of the bath \footnote{As noted in \cite{hb2}, if we consider a non-Boltzmannian distribution $f({\bf v})$ in $d>1$, we find that the Fokker-Planck equation does {\it not} relax towards $f({\bf v})$ but to another distribution function that is also non-Boltzmannian. However, as we have indicated, this approach is not self-consistent because $f({\bf v})$ is not steady on the timescale $Nt_D$ and cannot therefore form a ``bath''.} on a typical time
\begin{equation}
t_{R}^{bath}\sim N t_D.
\label{rbath1}
\end{equation}
This is the timescale controlling the relaxation of the test particle, i.e. the time needed by the test particle to acquire the distribution of the bath (for 3D stellar systems and plasmas, there are logarithmic corrections, so that $t_{R}^{bath}\sim (N/\ln N)t_D$ and $t_{R}^{bath}\sim (\Lambda/\ln \Lambda)t_D$ respectively). The timescale $t_{R}^{bath}$ should not be confused with the timescale $t_R$ controlling the relaxation of the system as a whole (see Sec. \ref{sec_relaxwhole}). In $d>1$ they have the same scaling but in $d=1$ they are totally different since $t_R>Nt_D$ (for spatially homogeneous systems). For one dimensional plasmas, the difference in behavior of  ``distinguished'' particles that relaxes on a timescale $\Lambda t_D$ and the overall population that relax on a timescale $\Lambda^2 t_D$  is shown in \cite{feix,rouetfeix}. As a clear sign of the inequivalence of these two descriptions (evolution of the system as a whole and relaxation of a test particle in a bath), we note that the Lenard-Balescu equation (\ref{lb27}) conserves the energy while the Fokker-Planck equation (\ref{conn5}) does not.

\section{Thermal bath  in $d$ dimensions}
\label{sec_tb}

\subsection{Einstein relation}
\label{sec_einstein}

We consider a thermal bath in $d$ dimensions that is formed by field particles at statistical equilibrium having the isothermal (Maxwell-Boltzmann) distribution
\begin{eqnarray}
f({\bf v})= \left (\frac{\beta m}{2\pi}\right )^{d/2} \rho\, e^{-\frac{1}{2}\beta m v^2}. \label{tb1}
\end{eqnarray}
Substituting the identity
\begin{eqnarray}
\frac{\partial f}{\partial {\bf v}}=-\beta m f({\bf v}){\bf v},
\label{tb2}
\end{eqnarray}
in Eq. (\ref{ess14again}), we obtain
\begin{eqnarray}
{F}_i^{pol}=-\pi (2\pi)^d m^2\beta\int d{\bf k}\, d{\bf v}'\, k_i \frac{\hat{u}(k)^2}{|\epsilon({\bf k},{\bf k}\cdot {\bf v})|^2}\delta\lbrack {\bf k}\cdot ({\bf v}- {\bf v}')\rbrack ({\bf k}\cdot {\bf v}') f({\bf v}').
\label{tb3}
\end{eqnarray}
Using the $\delta$-function to replace ${\bf k}\cdot {\bf v}'$ by ${\bf k}\cdot {\bf v}$,  and comparing the resulting expression with Eq. (\ref{diff7again}), we find that
\begin{eqnarray}
{\bf F}_i^{pol}=-D_{ij}({\bf v})\beta m v_j=-D_{\|}(v) \beta m v_i, \label{tb4}
\end{eqnarray}
where $D_{\|}(v)$ is the diffusion coefficient in the direction parallel to the velocity ${\bf v}$ of the test particles (see Eqs. (\ref{laa4}) and (\ref{laa8}) below that are valid for any isotropic distribution of the bath). The friction coefficient $\xi(v)=D_{\|}(v)\beta m$ is given by an Einstein relation expressing the fluctuation-dissipation theorem. We stress that the Einstein relation is valid for the friction due to the polarization ${\bf F}_{pol}$, not for the total (dynamical) friction  ${\bf A}$ that has a more complex expression due to the term $\partial_j D_{ij}$.  We do not have this subtlety for the usual Brownian motion for which the diffusion coefficient is constant. For a thermal bath, using
Eq. (\ref{tb4}), the Fokker-Planck equation (\ref{fp8}) takes the
form
\begin{equation}
\label{tb6}\frac{\partial P}{\partial t}=\frac{\partial}{\partial v_{i}} \left \lbrack D_{ij}({\bf v})\left (\frac{\partial P}{\partial
v_{j}}+\beta m  P v_j\right )\right\rbrack,
\end{equation}
where $D_{ij}({\bf v})$ is given by Eq. (\ref{diff7again}) with
Eq. (\ref{tb1}). This equation is similar to
the Kramers equation in Brownian theory \cite{risken}. However, in the present case, the diffusion coefficient is
anisotropic and depends on the velocity of the test particle. For $t\rightarrow +\infty$, the distribution of the test particle relaxes towards the Maxwell-Boltzmann distribution $P_e({\bf v})=(\beta m/2\pi)^{d/2}e^{-\beta m v^2/2}$. As we have seen, the relaxation time scales like $t_{R}^{bath}\sim N t_D$.

\subsection{General expression of the diffusion coefficient}
\label{sec_tp}

By introducing the representation
\begin{equation}
\label{isob3} \delta\lbrack {\bf k}\cdot ({\bf v}-{\bf v}')\rbrack=\int_{-\infty}^{+\infty} e^{i{\bf k}\cdot ({\bf v}-{\bf v}')t}\, {dt\over 2\pi},
\end{equation}
for the delta function in Eq. (\ref{diff7again}), the diffusion tensor can be rewritten
\begin{equation}
\label{isob4} D_{ij}({\bf v})= (2\pi)^{2d}m \int_{0}^{+\infty} dt \int d{\bf k}\, {k_i k_j}{\hat{u}({k})^{2}\over |\epsilon({\bf k},{\bf k}\cdot {\bf v})|^{2}}e^{i{\bf k}\cdot {\bf v}t}\hat{f}({\bf k}\,t).
\end{equation}
This equation can be directly obtained from the Kubo formula (\ref{diff5}). This shows that the auxiliary integration variable $t$ in Eq. (\ref{isob4}) actually represents the time. Since the distribution function (\ref{tb1}) is Gaussian, its three dimensional Fourier transform $\hat{f}({\bf k}\, t)$ is also Gaussian. The integration on time $t$, which is the one dimensional Fourier transform of a Gaussian, is therefore easily performed and we get
\begin{equation}
\label{isob5} D_{ij}({\bf v})=\pi (2\pi)^{d} \biggl ({\beta m\over 2\pi}\biggr )^{1/2}\, \rho m \int d{\bf k}\, {k_i k_j\over k}{\hat{u}({k})^{2}\over |\epsilon({\bf k},{\bf k}\cdot {\bf v})|^{2}}e^{-\beta m {({\bf k}\cdot {\bf v})^{2}\over 2k^{2}}}.
\end{equation}
Alternatively, this expression can be directly obtained from
Eq. (\ref{diff7again}) by introducing a cartesian system of
coordinates for ${\bf v}'$ with the $z$ axis taken along the direction
of ${\bf k}$ and performing the integration. On the other hand, for
the isothermal distribution (\ref{tb1}), the dielectric function is
given by Eq. (\ref{disp3}). Using Eq. (\ref{disp5}) and noting that
${\bf k}\cdot {\bf v}$ is real, we can rewrite the diffusion tensor in
the form
\begin{equation}
\label{isob9} D_{ij}({\bf v})=\int d\hat{\bf k}\, \hat{k}_{i}\hat{k}_{j}e^{-\frac{1}{2}\beta m (\hat{\bf k}\cdot {\bf v})^{2}} {\cal D}_d\left (\sqrt{\frac{\beta m}{2}}\, \hat{\bf k}\cdot {\bf v}\right ),
\end{equation}
where
\begin{equation}
\label{isob9b}
{\cal D}_d(x)=\frac{1}{2(2\pi)^{d-\frac{1}{2}}}\frac{v_m^3}{n}\int_0^{+\infty}  \frac{\eta(k)^2k^d}{\left\lbrack 1-\eta(k) B(x)\right\rbrack^{2}+\eta(k)^2C(x)^{2}}\, d{k}.
\end{equation}
In the foregoing formulae, we have defined $\hat{\bf k}={\bf k}/k$,
$B(x)=1-2x e^{-x^{2}}\int_{0}^{x}e^{y^{2}}dy$, $C(x)=\sqrt{\pi}x
e^{-x^2}$ and $\eta(k)=-(2\pi)^{d}\hat{u}(k)\beta m\rho$. In addition,
$v_{m}=({1/\beta m})^{1/2}$ is the r.m.s. velocity in one direction
and $n=\rho/m$ the numerical density of particles. The function $B(x)$
can be written $B(x)=1-2x {D}(x)$ where
${D}(x)=e^{-x^{2}}\int_{0}^{x}e^{y^{2}}dy$ is Dawson's integral. We
note the asymptotic behaviors $B(x)=1-2x^{2}+...$ for $x\rightarrow 0$
and $B(x)\sim -1/(2x^{2})$ for $x\rightarrow \pm\infty$. This function
vanishes at $x=0.92414...$ and achieves a minimum at
$(1.502,-0.28475)$. Finally, we note that $B(1)=-0.076159...$ and
$C(1)=0.652049...$.

Equations (\ref{isob9})-(\ref{isob9b}) provide the general form of the diffusion tensor of a test particle
in a thermal bath. The relaxation time can be estimated by $t_R^{bath}\sim v_m^2/D$. Since the velocity of the test particles scales like $\langle v^2\rangle \sim D t$, the relaxation time represents the time needed by the particle to acquire the typical velocity $v_m$ of the bath. Alternatively, the relaxation time can be estimated by the inverse of the friction coefficient $t_R^{bath}\sim 1/\xi$. Using the Einstein relation $\xi\sim D\beta m$, we again obtain $t_R^{bath}\sim v_m^2/D$. Then, using $D\sim v_m^3/nR^{d+1}$, which can be deduced from Eq. (\ref{isob9b}), and introducing the dynamical time $t_D^{bath}\sim R/v_m$, we obtain the scaling $t_R^{bath}\sim N t_D$. A more precise discussion of the relaxation time will be given in Paper II.

The foregoing expressions can be simplified for one-dimensional systems. In $d=1$, the Fokker-Planck equation takes the form
\begin{equation}
\label{casp1}
\frac{\partial P}{\partial t}=\frac{\partial}{\partial v} \left \lbrack D(v)\left (\frac{\partial P}{\partial
v}+\beta m  P v\right )\right\rbrack,
\end{equation}
where the diffusion coefficient is given by
\begin{equation}
\label{casp2} D(v)=2e^{-\frac{1}{2}\beta m v^{2}} {\cal D}_1\left (\sqrt{\frac{\beta m}{2}}\, v\right ),
\end{equation}
with
\begin{equation}
\label{casp3}
{\cal D}_1(x)=\frac{1}{2(2\pi)^{\frac{1}{2}}}\frac{v_m^3}{n}\int_0^{+\infty}  \frac{\eta(k)^2k}{\left\lbrack 1-\eta(k) B(x)\right\rbrack^{2}+\eta(k)^2C(x)^{2}}\, d{k}.
\end{equation}

\subsection{Neglect of collective effects: Landau approximation} 
\label{sec_laa}

If we neglect collective effects, which amounts to taking  $|\epsilon({\bf k},{\bf k}\cdot {\bf v})|= 1$ or, equivalently, $B=C=0$ in the equations of Sect. \ref{sec_tp}, we can write the diffusion tensor as
\begin{equation}
\label{lan1} D_{ij}({\bf v})={\cal D}_d\int d\hat{\bf k}\, \hat{k}_{i}\hat{k}_{j}e^{-\frac{1}{2}\beta m (\hat{\bf k}\cdot {\bf v})^{2}},
\end{equation}
with
\begin{equation}
\label{lan2}
{\cal D}_d=\frac{1}{2(2\pi)^{d-\frac{1}{2}}}\frac{v_m^3}{n}\int_0^{+\infty}  \eta(k)^2{k^d}\, d{k}.
\end{equation}
In the Landau approximation, the potential of interaction only appears in a multiplicative constant ${\cal D}_d$ that fixes the relaxation time. Therefore, general expressions of the diffusion tensor can be given \cite{landaud}.

In $d=3$, introducing a spherical system of coordinates, we obtain
\begin{eqnarray}
D_{ij}({v})=\left \lbrack D_{\|}(v)-{1\over 2}D_{\perp}(v)\right \rbrack {v_i v_j\over v^{2}}+{1\over 2}D_{\perp}(v)\delta_{ij},
\label{laa4}
\end{eqnarray}
with
\begin{eqnarray}
D_{\|}(v)=2\pi^{3/2}{\cal D}_3 \frac{G\left (\sqrt{\frac{\beta m}{2}}v\right )}{\sqrt{\frac{\beta m}{2}}v},\qquad
D_{\perp}(v)=2\pi^{3/2}{\cal D}_3 \frac{{\rm erf}\left (\sqrt{\frac{\beta m}{2}}v\right )-G\left (\sqrt{\frac{\beta m}{2}}v\right )}{\sqrt{\frac{\beta m}{2}}v},
\label{laa5}
\end{eqnarray}
\begin{eqnarray}
G(w)={2\over\sqrt{\pi}}{1\over w^{2}}\int_{0}^{w}t^{2}e^{-t^{2}}dt={1\over 2w^{2}}\biggl\lbrack {\rm erf}(w)-{2w\over \sqrt{\pi}}e^{-w^{2}}\biggr\rbrack,
\label{laa6}
\end{eqnarray}
where ${\rm erf}(w)$ is the error function. In the above expressions, $D_{\|}(v)$ and $D_{\perp}(v)$ are the diffusion
coefficients in the directions parallel and perpendicular to the
velocity of the test particle. We note the asymptotic behaviors
\begin{eqnarray}
D_{\|}(0)=\frac{4\pi{\cal D}_3}{3},\quad D_{\perp}(0)=\frac{8\pi{\cal D}_3}{3},\quad D_{\|}(v)\sim_{+\infty}\frac{\pi^{3/2}{\cal D}_3}{\left (\sqrt{\frac{\beta m}{2}}v\right )^3},\quad D_{\perp}(v)\sim_{+\infty}\frac{2\pi^{3/2}{\cal D}_3}{\sqrt{\frac{\beta m}{2}}v}.
\label{ass1}
\end{eqnarray}

In $d=2$, introducing a polar system of coordinates, we obtain
\begin{eqnarray}
D_{ij}({v})=\left \lbrack D_{\|}(v)-D_{\perp}(v)\right\rbrack {v_i v_j \over v^{2}}+D_{\perp}(v)\delta_{ij},
\label{laa8}
\end{eqnarray}
with
\begin{eqnarray}
D_{\|}(v)={\cal D}_2 \pi e^{-\frac{\beta m}{4}v^2}\biggl\lbrack I_{0}\biggl (\frac{\beta m}{4}v^2\biggr )- I_{1}\biggl (\frac{\beta m}{4}v^2\biggr )\biggr\rbrack, \qquad
D_{\perp}(v)= {\cal D}_2 \pi e^{-\frac{\beta m}{4}v^2}\biggl\lbrack I_{0}\biggl (\frac{\beta m}{4}v^2\biggr )+I_{1}\biggl (\frac{\beta m}{4}v^2\biggr )\biggr\rbrack,
\label{laa9}
\end{eqnarray}
where $I_n(x)$ are the modified Bessel functions.  We note the asymptotic behaviors
\begin{eqnarray}
D_{\|}(0)=\pi{\cal D}_2,\quad D_{\perp}(0)=\pi{\cal D}_2,\quad D_{\|}(v)\sim_{+\infty}\frac{\sqrt{\pi}{\cal D}_2}{\left (\sqrt{\frac{\beta m}{2}}v\right )^3},\quad D_{\perp}(v)\sim_{+\infty}\frac{2\sqrt{\pi}{\cal D}_2}{\sqrt{\frac{\beta m}{2}}v}.
\label{ass2}
\end{eqnarray}

Finally, in $d=1$,
\begin{eqnarray}
D(v)=2{\cal D}_1 e^{-\frac{1}{2}m\beta v^2}.
\label{laa10}
\end{eqnarray}

\subsection{Asymptotic behaviors and Debye-H\"uckel approximation}
\label{sec_asyb}

In this subsection, we take collective effects into account but consider asymptotic expressions of the diffusion coefficient.

For small velocities $v\rightarrow 0$, the function defined by Eq. (\ref{isob9b}) can be approximated by
\begin{equation}
\label{asyb1}
{\cal D}_d(0)=\frac{1}{2(2\pi)^{d-\frac{1}{2}}}\frac{v_m^3}{n}\int_0^{+\infty}  \frac{\eta(k)^2k^d}{\left\lbrack 1-\eta(k)\right\rbrack^{2}}\, d{k}.
\end{equation}
Therefore, when $v\rightarrow 0$, the diffusion coefficient is given by Eq. (\ref{lan1}) with ${\cal D}_d$ replaced by ${\cal D}_d(0)$. If we come back to the expression (\ref{isob5}) of the diffusion coefficient, and use Eq. (\ref{disp3}), this amounts to replacing the dressed potential of interaction $\hat{u}_{dressed}(k)={\hat{u}(k)}/{|\epsilon({\bf k},{\bf k}\cdot {\bf v})|}$ by
\begin{equation}
\label{asyb2}
\hat{u}_{DH}(k)=\frac{\hat{u}(k)}{1+(2\pi)^d\hat{u}(k)\beta m\rho}.
\end{equation}
When this formula is specialized to the Coulombian potential  $(2\pi)^d\hat{u}(k)=S_d e^2/m^2 k^2$, we see that Eq. (\ref{asyb2}) is the Fourier transform of the Debye-H\"uckel potential
\begin{equation}
\label{asyb2b}
(2\pi)^d\hat{u}_{DH}(k)=\frac{S_d e^2}{m^2}\frac{1}{k^2+k_D^2}.
\end{equation} 
The replacement of the Coulombian potential by the  Debye-H\"uckel potential is a simple way to take into account collective effects in the kinetic theory by using the {\it static} results on screening. This procedure has been introduced phenomenologically by several authors in plasma physics before Lenard and Balescu provided the rigorous solution (see Paper II). The present considerations show that the Debye-H\"uckel approximation is valid only for a thermal bath and for small velocities of the test particle. The results of Lenard and Balescu are more general because they account for {\it dynamical} screening and take the deformation of the polarization cloud, due to the motion of the test particle, into account.

We now consider the case of large velocities $v\rightarrow +\infty$. In $d=3$ and $d=2$ dimensions, it is not difficult to see that in the evaluation of $D_{\perp}(v)$, the function ${\cal D}_d(x)$ can be replaced by ${\cal D}_d(0)$, while in the evaluation of $D_{\|}(v)$, it can be replaced by
\begin{equation}
\label{asyb3}
{\cal D}_d(1)=\frac{1}{2(2\pi)^{d-\frac{1}{2}}}\frac{v_m^3}{n}\int_0^{+\infty}  \frac{\eta(k)^2k^d}{\left\lbrack 1-\eta(k) B(1)\right\rbrack^{2}+\eta(k)^2C(1)^{2}}\, d{k}.
\end{equation}
As a result, for $v\rightarrow +\infty$, the diffusion coefficient $D_{\perp}(v)$ is given by Eq. (\ref{lan1}) with ${\cal D}_d$ replaced by ${\cal D}_d(0)$, and the diffusion coefficient $D_{\|}(v)$ is given by Eq. (\ref{lan1}) with  ${\cal D}_d$ replaced by ${\cal D}_d(1)$. By contrast, in $d=1$ dimension, there is no general expression for the large velocity behavior of the diffusion coefficient. It depends on the precise form of the potential of interaction. Analytical expressions will be given in Paper II for particular systems.

\section{Out-of-equilibrium bath in $d=1$}
\label{sec_nontb}

In one dimension, we have seen that the Lenard-Balescu collision term vanishes due to the absence of resonance. Therefore, a stable steady state $f(v)$ of the Vlasov equation does not evolve under the effect of collisions on a timescale of order $Nt_D$. We can therefore develop a test particle approach in a bath characterized by an arbitrary Vlasov stable distribution $f(v)$, not necessarily the Maxwell-Boltzmann distribution of statistical equilibrium.  Since the distribution $f(v)$ does not change on the timescale $t_R^{bath}\sim Nt_D$ corresponding to the relaxation of the test particle, we can consider that $f(v)$ is ``frozen'' on this timescale, so the bath approach is self-consistent.

In $d=1$, the Fokker-Planck equation (\ref{conn5}) takes the form
\begin{equation}
\label{genn1}  {\partial P\over\partial t}={\partial\over\partial v}\biggl\lbrack D(v)\biggl ({\partial P\over\partial v}-P {d\ln f\over dv}\biggr )\biggr\rbrack,
\end{equation}
with a diffusion coefficient
\begin{equation}
\label{genn2} D(v)=m f(v)\int_{0}^{+\infty} dk {4\pi^{2}\hat{u}(k)^{2}k\over |\epsilon(k,kv)|^{2}}.
\end{equation}
According to Eq. (\ref{pln}) of Appendix \ref{sec_disp}, one has
\begin{equation}
\label{genn3}
|\epsilon(k,kv)|^{2}=\left\lbrack 1-2\pi \hat{u}(k) {\cal P}\int_{-\infty}^{+\infty}\frac{f'(u)}{u-v}\, du\right\rbrack^2+4\pi^4 \hat{u}(k)^2f'(v)^2.
\end{equation}
For the isothermal distribution (\ref{tb1}), we recover the results of Section \ref{sec_tb}. For the waterbag distribution, using Eq. (\ref{disp2}), we obtain
\begin{equation}
\label{genn4}
D(v)= \frac{\rho m}{2v_m}\int_{0}^{+\infty} dk {4\pi^2\hat{u}(k)^{2}k\over \left\lbrack 1+\frac{2\pi\hat{u}(k)\rho}{v_m^2-v^2}\right\rbrack^2},
\end{equation}
for $-v_m\le v\le v_m$ and $D(v)=0$ otherwise.  When collective
effects are neglected, the diffusion coefficient reduces to the form
\begin{equation}
\label{genn2b} D(v)=m f(v)\int_{0}^{+\infty}  4\pi^{2}\hat{u}(k)^{2} k\, dk.
\end{equation}
It is proportional to the distribution function $f(v)$.

Equation (\ref{genn1}) is similar to a Fokker-Planck equation
describing the motion of a Brownian particle in a potential $U({v})=-\ln f({v})$ created by the field particles. The distribution function of the test particle $P(v,t)$
relaxes towards the distribution of the bath $f(v)$ for
$t\rightarrow +\infty$. The typical timescale governing the approach
of the test particle to the bath distribution is $t_R^{bath}\sim N t_D$. The fact that the distribution $P(v,t)$ of a test particle relaxes towards {\it any} stable distribution $f(v)$ of the field particles for $d=1$ explains why the distribution of the field particles  does not
change on a timescale $\sim Nt_{D}$. The situation is different in $d>1$ \cite{hb2}.

\section{The effect of an external stochastic forcing}
\label{sec_forcing}

In a recent paper, Nardini {\it et al.} \cite{nardini} have studied
long-range interacting systems perturbed by external stochastic forces
and derived a nonlinear Fokker-Planck equation describing the
evolution of the system. We have independently considered the same
problem. However, our approach is different and leads to a different
final equation. The reason of this difference is unknown to us but we
think that it is interesting to present our calculation in detail. We
therefore consider the case of a system with long-range interactions
that is submitted to an external stochastic potential $\Phi_e({\bf
r},t)$. The Klimontovich equation (\ref{pvm9}) remains valid provided
that we make the substitution $\Phi_d\rightarrow
\Phi_d+\Phi_e$. We shall treat the stochastic potential $\Phi_e({\bf
r},t)$ as a small perturbation. We shall also assume that the system
remains spatially homogeneous during the whole evolution. Then,
decomposing $f_d=f+\delta f$ and $\Phi_d=\Phi+\delta\Phi$ into a
smooth component plus a fluctuating component, and making a
quasilinear approximation as in Section \ref{sec_lb}, we find that
Eqs. (\ref{lb3}) and (\ref{lb4}) are replaced by
\begin{equation}
\frac{\partial f}{\partial t}=\frac{\partial}{\partial {\bf v}}\cdot \left\langle \delta f \nabla\delta\Phi\right\rangle+\frac{\partial}{\partial {\bf v}}\cdot \left\langle \delta f \nabla\Phi_e\right\rangle,
\label{for1}
\end{equation}
\begin{equation}
\frac{\partial\delta f}{\partial t}+{\bf v}\cdot \frac{\partial \delta f}{\partial {\bf r}}-\nabla\delta\Phi\cdot \frac{\partial f}{\partial {\bf v}}-\nabla\Phi_e\cdot \frac{\partial f}{\partial {\bf v}}=0.
\label{for2}
\end{equation}
Equation (\ref{for1}) for the smooth component is exact while, in writing Eq. (\ref{for2}) for the fluctuation,  we have neglected the nonlinear terms $\delta f\delta\Phi-\langle \delta f\delta\Phi\rangle$ and $\delta f \Phi_e-\langle\delta f \Phi_e\rangle$. If we take the  Fourier-Laplace transform of Eq. (\ref{for2}), we find that
\begin{equation}
\delta\tilde f ({\bf k},{\bf v},\omega)=\frac{{\bf k}\cdot \frac{\partial f}{\partial {\bf v}}}{{\bf k}\cdot {\bf v}-\omega}\left\lbrack \delta\tilde\Phi({\bf k},\omega)+\tilde\Phi_e({\bf k},\omega)\right\rbrack+\frac{\delta\hat f({\bf k},{\bf v},0)}{i({\bf k}\cdot {\bf v}-\omega)}.
\label{for3}
\end{equation}
Substituting Eq. (\ref{for3}) in Eq. (\ref{lb10}), we find that the Fourier-Laplace transform of the fluctuations of the potential is related to the initial condition and to the stochastic potential by
\begin{equation}
\delta\tilde\Phi({\bf k},\omega)=(2\pi)^d\frac{\hat{u}(k)}{\epsilon({\bf k},\omega)}\int d{\bf v}\, \frac{\delta\hat f({\bf k},{\bf v},0)}{i({\bf k}\cdot {\bf v}-\omega)}+\frac{1-\epsilon({\bf k},\omega)}{\epsilon({\bf k},\omega)}\tilde\Phi_e({\bf k},\omega).
\label{for4}
\end{equation}

We can use these expressions to compute the terms appearing on the right hand side of Eq. (\ref{for1}). The first term can be written
\begin{equation}
\left\langle\delta f \nabla\delta\Phi\right\rangle=\int d{\bf k}\int_{\cal C} \frac{d\omega}{2\pi}\int d{\bf k'}\int_{\cal C} \frac{d\omega'}{2\pi} \, i {\bf k}' e^{i({\bf k}\cdot {\bf r}-\omega t)}e^{i({\bf k}'\cdot {\bf r}-\omega' t)}\langle \delta\tilde f({\bf k},{\bf v},\omega)\delta\tilde\Phi({\bf k}',\omega')\rangle.
\label{for5}
\end{equation}
Using Eq. (\ref{for3}), we obtain
\begin{equation}
\langle \delta\tilde f({\bf k},{\bf v},\omega)\delta\tilde\Phi({\bf k}',\omega')\rangle=\frac{{\bf k}\cdot \frac{\partial f}{\partial {\bf v}}}{{\bf k}\cdot {\bf v}-\omega}\left\lbrack \langle \delta\tilde\Phi({\bf k},\omega)\delta\tilde\Phi({\bf k}',\omega')\rangle+\langle \tilde\Phi_e({\bf k},\omega)\delta\tilde\Phi({\bf k}',\omega')\rangle\right\rbrack+\frac{\langle \delta\hat f({\bf k},{\bf v},0)\delta\tilde\Phi({\bf k}',\omega')\rangle}{i({\bf k}\cdot {\bf v}-\omega)}.
\label{for6}
\end{equation}
Some of the terms appearing in this expression have already been computed in Sect. \ref{sec_lb}. They lead to the Lenard-Balescu equation (\ref{lb27}). In this section, we shall only write down the new term arising from the stochastic forcing. Using Eq. (\ref{for4}), this term is found to be
\begin{equation}
\langle \delta\tilde f({\bf k},{\bf v},\omega)\delta\tilde\Phi({\bf k}',\omega')\rangle=\frac{{\bf k}\cdot \frac{\partial f}{\partial {\bf v}}}{{\bf k}\cdot {\bf v}-\omega}\frac{1-\epsilon({\bf k}',\omega')}{\epsilon({\bf k},\omega)\epsilon({\bf k}',\omega')} \langle \tilde\Phi_e({\bf k},\omega)\tilde\Phi_e({\bf k}',\omega')\rangle.
\label{for7}
\end{equation}
We assume that the stochastic potential $\Phi_e({\bf r},t)$ is $\delta$-correlated in time and that the spatial correlator only depends on the distance ${\bf r}-{\bf r}'$. Under these assumptions, we have
\begin{equation}
\langle\tilde\Phi_e({\bf k},\omega)\tilde\Phi_e({\bf k}',\omega')\rangle=P(k)\delta({\bf k}+{\bf k}')\delta(\omega+\omega'),
\label{for8}
\end{equation}
where $P(k)$ is the amplitude of the noise in Fourier space.  Equation (\ref{for7}) can then be rewritten
\begin{equation}
\langle \delta\tilde f({\bf k},{\bf v},\omega)\delta\tilde\Phi({\bf k}',\omega')\rangle=\frac{{\bf k}\cdot \frac{\partial f}{\partial {\bf v}}}{{\bf k}\cdot {\bf v}-\omega}\frac{1-\epsilon({\bf k},\omega)^*}{|\epsilon({\bf k},\omega)|^2} P(k)\delta({\bf k}+{\bf k}')\delta(\omega+\omega').
\label{for8b}
\end{equation}
The new contribution of the first term in the right hand side of Eq. (\ref{for1}) arising from the stochastic forcing is therefore
\begin{equation}
\left\langle\delta f \nabla\delta\Phi\right\rangle_i=-\frac{i}{2\pi} \int d{\bf k}\int\frac{d\omega}{2\pi} \, {k}_i \frac{{\bf k}\cdot \frac{\partial f}{\partial {\bf v}}}{{\bf k}\cdot {\bf v}-\omega} \frac{1-\epsilon({\bf k},\omega)^*}{|\epsilon({\bf k},\omega)|^2}P(k).
\label{for9}
\end{equation}
Proceeding similarly, we find that the second term in  Eq. (\ref{for1}) can be written
\begin{equation}
\left\langle\delta f \nabla\Phi_e\right\rangle_i=-\frac{i}{2\pi} \int d{\bf k}\int\frac{d\omega}{2\pi} \, {k}_i \frac{{\bf k}\cdot \frac{\partial f}{\partial {\bf v}}}{{\bf k}\cdot {\bf v}-\omega} \frac{\epsilon({\bf k},\omega)^*}{|\epsilon({\bf k},\omega)|^2}P(k).
\label{for10}
\end{equation}
Therefore, the net effect of the  stochastic forcing is to lead to an anisotropic diffusion of the form
\begin{equation}
\left (\frac{\partial f}{\partial t}\right )_{forcing}=\frac{\partial}{\partial v_i}\left(D_{ij}^{forcing}[f] \frac{\partial f}{\partial {v}_j}\right ),
\label{for11}
\end{equation}
with a diffusion tensor $D_{ij}^{forcing}=D_{ij}^{I}+D_{ij}^{II}$, where
\begin{equation}
D_{ij}^{I}[f]=-\frac{i}{2\pi}\int d{\bf k}\int\frac{d\omega}{2\pi} \, {k}_i k_j \frac{1}{{\bf k}\cdot {\bf v}-\omega}\frac{1-\epsilon({\bf k},\omega)^*}{|\epsilon({\bf k},\omega)|^2}P(k),
\label{for12}
\end{equation}
\begin{equation}
D_{ij}^{II}[f]=-\frac{i}{2\pi}\int d{\bf k}\int\frac{d\omega}{2\pi} \, {k}_i k_j \frac{1}{{\bf k}\cdot {\bf v}-\omega}\frac{\epsilon({\bf k},\omega)^*}{|\epsilon({\bf k},\omega)|^2}P(k).
\label{for13}
\end{equation}
The total diffusion tensor is
\begin{equation}
D_{ij}^{forcing}[f]=-\frac{i}{2\pi}\int d{\bf k}\int\frac{d\omega}{2\pi} \, {k}_i k_j \frac{1}{{\bf k}\cdot {\bf v}-\omega}\frac{1}{|\epsilon({\bf k},\omega)|^2}P(k).
\label{for14}
\end{equation}
Using the Landau prescription $\omega\rightarrow \omega+i0^+$ and the Plemelj formula (\ref{lb21}), we can replace $1/({\bf k}\cdot {\bf v}-\omega-i0^+)$ by $+i\pi\delta({\bf k}\cdot {\bf v}-\omega)$. Then, integrating over  $\omega$,
we obtain
\begin{equation}
D_{ij}^{forcing}[f]=\frac{1}{4\pi}\int d{\bf k} \, {k}_i k_j \frac{P(k)}{|\epsilon({\bf k},{\bf k}\cdot {\bf v})|^2}.
\label{for15}
\end{equation}
We note that,  because of the presence of the dielectric function $\epsilon({\bf
k},{\bf k}\cdot {\bf v})$ defined by Eq. (\ref{lb12}) in the denominator of Eq. (\ref{for15}), $D_{ij}^{forcing}[f]$ is a {\it functional} of $f({\bf v},t)$. 
For non-interacting particles, $\hat{u}({\bf k})=0$, the diffusion coefficient arising from the forcing is given by Eq. (\ref{for15}) with $\epsilon({\bf k},{\bf
k}\cdot {\bf v})=1$. This is just a constant depending on the amplitude
$P(k)$ of the noise. For interacting particles, the amplitude $P(k)$ is ``dressed'' by the polarizaion cloud encapsulated in the dielectric function. The part $D_{ij}^{I}$ of the diffusion tensor is explicitly computed in Appendix \ref{sec_diffun}.

Because of the stochastic force, the distribution function $f({\bf v},t)$ will diffuse and will not reach a steady state. However, a steady state can be reached if we assume that the particles, in addition to the stochastic force, also experience a friction force $-\xi {\bf v}$ due to the presence of an inert medium. If the friction force is small so that its influence can be neglected in Eq. (\ref{for2}), the calculation of the diffusion coefficient is not modified and we readily obtain a nonlinear Fokker-Planck equation of the form
\begin{equation}
\left (\frac{\partial f}{\partial t}\right )_{forcing + dissipation}=\frac{\partial}{\partial v_i}\left(D_{ij}^{forcing}[f] \frac{\partial f}{\partial {v}_j}+\xi f v_i\right ).
\label{for16fric}
\end{equation}
This Fokker-Planck equation is expected to relaxe, on a timescale of the order $\xi^{-1}$, towards a non-Boltzmannian steady state determined, in a self-consistent manner, by the diffusion tensor $D_{ij}^{forcing}[f]$. In fact, since this equation is nonlinear, it is not clear whether an equilibrium state will be finally achieved.

Finally, if we develop a test particle approach and make a ``bath approximation'', we can compute the diffusion tensor as in Section \ref{sec_stoch}. In that case, we find that the new term in the Fokker-Planck equation (\ref{conn5}) due to the forcing and the dissipation is
\begin{equation}
\left (\frac{\partial P}{\partial t}\right )_{forcing + dissipation}=\frac{\partial}{\partial v_i}\left(D_{ij}^{forcing}({\bf v}) \frac{\partial P}{\partial {v}_j}+\xi P v_i\right ),
\label{for16}
\end{equation}
where now $f({\bf v})$ is the time-independent distribution of the bath. In particular, for a thermal bath, using the expression (\ref{disp3}) of the dielectric function, we get
\begin{equation}
D_{ij}^{forcing}({\bf v})=\frac{1}{4\pi}\int  \frac{{k}_i k_j P(k)}{\left\lbrack 1-\eta(k)B\left (\sqrt{\frac{\beta m}{2}}\frac{{\bf k}\cdot {\bf v}}{k}\right )\right\rbrack^2+\eta(k)^2 C\left (\sqrt{\frac{\beta m}{2}}\frac{{\bf k}\cdot {\bf v}}{k}\right )^2}\, d{\bf k}.
\label{for15b}
\end{equation}
This Fokker-Planck equation relaxes, on a timescale of the order $\xi^{-1}$, towards a non-Boltzmannian steady state determined by the diffusion tensor $D_{ij}^{forcing}({\bf v})$ that is itself determined by the distribution of the bath $f({\bf v})$. Simplified expressions will be given in Paper II. The results of this section can also be generalized to spatially inhomogeneous systems, using angle-action variables as in \cite{newangleaction}.

{\it Remark:} We emphasize that the effect of an external forcing on the dynamics of systems with long-range interactions is physically different from considering Brownian particles with long-range interactions like in \cite{hb2,baldo1,baldo2}. There exists, however, qualitative analogies. For Brownian particles, the effect of the thermal bath is represented, in the kinetic equation, by a linear Fokker-Planck operator of the Kramers type. The thermal bath breaks the conservation of energy and the system ultimately  relaxes towards the Boltzmann distribution with the temperature of the bath, which is the canonical distribution. In the present case, the forcing and friction terms also break the conservation of energy (in a different manner) and the system ultimately relaxes towards a non-Boltzmannian distribution depending implicitly on the amplitude $P(k)$ of the stochastic force in Fourier space.

\section{Conclusion}
\label{sec_conclusion}

In this paper, we have reviewed and completed the existing literature
on the kinetic theory of systems with long-range interactions. The
Lenard-Balescu equation introduced in the context of plasma physics
has a wider scope, as it describes a large class of systems with
(attractive or repulsive) weak long-range interactions. We have
considered a test particle approach and given explicit expressions of
the diffusion coefficient and friction force arising in the
Fokker-Planck equation. We have stressed the analogies and the
differences between the Lenard-Balescu equation describing the
evolution of the system as a whole and the Fokker-Planck equation
describing the relaxation of a test particle in a bath.  We have also
considered the effect of a stochastic forcing on the evolution of a
system with long-range interactions and we have shown that it leads to
an anisotropic diffusion with a diffusion tensor expressed as a
functional of the distribution function itself.

The general results that we have presented in this paper will be used, in Paper II, to investigate the kinetic theory of particular systems with long-range interactions.

\appendix

\section{Auto-correlation of the fluctuations of the one-particle density}
\label{sec_auto}

According to Eq. (\ref{lb8}), we have
\begin{eqnarray}
\langle \delta\hat f({\bf k},{\bf v},0)\delta\hat f({\bf k}',{\bf v}',0)\rangle&=&\int\frac{d{\bf r}}{(2\pi)^d}\frac{d{\bf r}'}{(2\pi)^d}e^{-i({\bf k}\cdot {\bf r}+{\bf k}'\cdot {\bf r}')}\langle \delta f({\bf r},{\bf v},0)\delta f({\bf r}',{\bf v}',0)\rangle\nonumber\\
&=&\int\frac{d{\bf r}}{(2\pi)^d}\frac{d{\bf r}'}{(2\pi)^d}e^{-i({\bf k}\cdot {\bf r}+{\bf k}'\cdot {\bf r}')}\left\lbrack \langle f_d({\bf r},{\bf v},0)f_d({\bf r}',{\bf v},0)\rangle-f({\bf v})f({\bf v}')\right\rbrack.
\label{auto1}
\end{eqnarray}
The expression of the discrete distribution function recalled in the first paragraph of Sec. \ref{sec_lb} leads to
\begin{eqnarray}
\langle f_d({\bf r},{\bf v},0)f_d({\bf r}',{\bf v}',0)\rangle&=&m^2\sum_{i,j}\left\langle \delta({\bf r}-{\bf r}_i)\delta({\bf v}-{\bf v}_i)\delta({\bf r}'-{\bf r}_j)\delta({\bf v}'-{\bf v}_j)\right\rangle\nonumber\\
&=&m^2\sum_{i}\left\langle \delta({\bf r}-{\bf r}_i)\delta({\bf v}-{\bf v}_i)\delta({\bf r}-{\bf r}')\delta({\bf v}-{\bf v}')\right\rangle\nonumber\\
&+&m^2\sum_{i\neq j}\left\langle \delta({\bf r}-{\bf r}_i)\delta({\bf v}-{\bf v}_i)\delta({\bf r}'-{\bf r}_j)\delta({\bf v}'-{\bf v}_j)\right\rangle\nonumber\\
&=& m f({\bf v}) \delta({\bf r}-{\bf r}')\delta({\bf v}-{\bf v}')+f({\bf v})f({\bf v}'),
\label{auto2}
\end{eqnarray}
where we have assumed that, initially, there is no correlation between different particles (if there are correlations described by a smooth distribution $h_2({\bf r}-{\bf r}',{\bf v},{\bf v}')$, it can be shown that their contribution decays rapidly with time  \cite{pitaevskii} so  they have no effect on the final form of the collision term). Combining Eqs. (\ref{auto1}) and (\ref{auto2}), we obtain
\begin{equation}
\langle \delta\hat f({\bf k},{\bf v},0)\delta\hat f({\bf k}',{\bf v}',0)\rangle=\frac{1}{(2\pi)^d}\delta({\bf k}+{\bf k}')\delta({\bf v}-{\bf v}') m f({\bf v}).
\label{auto3}
\end{equation}

\section{Dispersion relation}
\label{sec_disp}

We consider the linear dynamical stability of a spatially homogeneous
steady state $f({\bf v})$ of the Vlasov equation
(\ref{pvm13}). Decomposing the perturbation in a sum of plane waves of
the form $\delta f\sim e^{i({\bf k}\cdot {\bf r}-\omega t)}$ and
substituting this decomposition in the linearized Vlasov equation
(\ref{lb4}) we obtain the dispersion relation $\epsilon({\bf
k},\omega)=0$, where the dielectric function is defined by
Eq. (\ref{lb12}). It determines the complex pulsation
$\omega=\omega_r+i\omega_i$ as a function of the wavenumber
$k$. Taking the $z$-axis in the direction of ${\bf k}$, we get
\begin{equation}
\label{disp1}
\epsilon(k,\omega)=1-(2\pi)^d\hat{u}(k)\int_{\cal C}\frac{f'(v)}{v-\frac{\omega}{k}}\, dv,
\end{equation}
where we have noted $v$ for $v_z$ and $f(v)$ for $\int f({\bf v})\, dv_x dv_y$. On the other hand, ${\cal C}$ is the Landau contour. Explicit expressions of the integral, depending on the sign of $\omega_i$, are given in \cite{cd}. In particular, when $\omega_i=0$, using the Plemelj formula (\ref{lb21}), we get
\begin{equation}
\label{pln}
\epsilon(k,\omega)=1-(2\pi)^d\hat{u}(k){\cal P}\int_{-\infty}^{+\infty}\frac{f'(v)}{v-\frac{\omega}{k}}\, dv-(2\pi)^d\hat{u}(k)i\pi f'\left (\frac{\omega}{k}\right).
\end{equation}
Let us first consider particular distributions.

$\bullet$ The {\it waterbag distribution} is defined by $f(v)=\rho/2v_m$ for $-v_m\le v\le v_m$ and $f(v)=0$ otherwise (we note that $v_m=0$ corresponds to $f(v)=\rho\delta(v)$). Using $f'(v)=(\rho/2v_m)\lbrack \delta(v+v_m)-\delta(v-v_m)\rbrack$, the dielectric function (\ref{disp1}) is explicitly  given by
\begin{equation}
\label{disp2}
\epsilon(k,\omega)=1+(2\pi)^d\hat{u}(k)\rho\frac{1}{v_m^2-(\omega/k)^2}.
\end{equation}
The dispersion relation can be written  $\omega^2=v_m^2k^2+(2\pi)^d\hat{u}(k)k^2\rho$. We note that $\omega$ is purely real or purely imaginary.

$\bullet$ For the {\it isothermal distribution} (\ref{tb1}),  introducing the notation $\eta(k)=-(2\pi)^{d}\hat{u}(k)\beta m\rho$, the dielectric function can be written
\begin{equation}
\label{disp3}
\epsilon(k,\omega)=1-\eta(k)W\left (\sqrt{\beta m}\frac{\omega}{k}\right ),
\end{equation}
where
\begin{equation}
\label{disp4}
W(z)=\frac{1}{\sqrt{2\pi}}\int_{\cal C}\frac{x}{x-z}e^{-\frac{x^2}{2}}\, dx,
\end{equation}
is the $W$-function of plasma physics \cite{ichimaru}. It has the explicit expression
\begin{equation}
\label{disp5}
W(z)=1-z e^{-\frac{z^2}{2}}\int_0^z e^{\frac{y^2}{2}}\, dy+i\sqrt{\frac{\pi}{2}}z e^{-\frac{z^2}{2}},
\end{equation}
for any $z\in \mathbf{C}$. We note that $W(0)=1$. On the other hand, for $z\in \mathbf{R}$, $W_r(z)=B(z/\sqrt{2})$ and $W_i(z)=C(z/\sqrt{2})$ where the functions $B(x)$ and $C(x)$ are defined in Section \ref{sec_tp}.

\section{The first component of the diffusion coefficient due to the forcing}
\label{sec_diffun}

The first component (\ref{for12}) of the diffusion coefficient due to the forcing can be written
\begin{equation}
D_{ij}^{I}[f]=\frac{i}{2\pi}\int d{\bf k}\int\frac{d\omega}{2\pi} \, {k}_i k_j \frac{1}{\omega-{\bf k}\cdot {\bf v}+i0^+}\frac{1-{\rm Re}\lbrack \epsilon({\bf k},\omega)\rbrack+i {\rm Im}\lbrack \epsilon({\bf k},\omega)\rbrack }{|\epsilon({\bf k},\omega)|^2}P(k),
\label{diffun1}
\end{equation}
Using the Plemelj formula (\ref{lb21}), we obtain
\begin{eqnarray}
D_{ij}^{I}[f]=\frac{1}{4\pi}\int d{\bf k}\, {k}_i k_j \frac{1-{\rm Re}\lbrack \epsilon({\bf k},{\bf k}\cdot {\bf v})\rbrack}{|\epsilon({\bf k},{\bf k}\cdot {\bf v})|^2}P(k)
-\frac{1}{4\pi^2}\int d{\bf k}\, {\cal P}\int d\omega \, {k}_i k_j \frac{1}{\omega-{\bf k}\cdot {\bf v}}\frac{{\rm Im}\lbrack \epsilon({\bf k},\omega)\rbrack }{|\epsilon({\bf k},\omega)|^2}P(k).
\label{diffun2}
\end{eqnarray}
On the other hand, the dielectric function (\ref{lb12}) can be written
\begin{equation}
\epsilon({\bf k},\omega)=1+(2\pi)^d\hat{u}(k){\cal P}\int  \frac{{\bf k}\cdot \frac{\partial f}{\partial {\bf v}}}{\omega-{\bf k}\cdot {\bf v}}\, d{\bf v}-i\pi (2\pi)^d\hat{u}(k)\int  {\bf k}\cdot \frac{\partial f}{\partial {\bf v}}\delta(\omega-{\bf k}\cdot {\bf v})\, d{\bf v}.
\label{diffun3}
\end{equation}
Substituting Eq. (\ref{diffun3}) in Eq. (\ref{diffun2}), we get
\begin{eqnarray}
D_{ij}^{I}[f]=\frac{(2\pi)^d}{4\pi}\int d{\bf k}\, {k}_i k_j \hat{u}(k) \, {\cal P}\int d{\bf v}' \frac{{\bf k}\cdot \frac{\partial f}{\partial {\bf v}'}}{{\bf k}\cdot ({\bf v}'-{\bf v})}\left (\frac{1}{|\epsilon({\bf k},{\bf k}\cdot {\bf v}')|^2}+\frac{1}{|\epsilon({\bf k},{\bf k}\cdot {\bf v})|^2}\right ) P(k).
\label{diffun4}
\end{eqnarray}


\begin{thebibliography}{}

\bibitem{houches}  {\small  {Dynamics and Thermodynamics of Systems with Long-Range Interactions}, edited by T. Dauxois, S. Ruffo, E. Arimondo and  M. Wilkens, Lectures  Notes in Physics {\bf 602} (Berlin: Springer, 2002)}
\bibitem{assise}  {\small {Dynamics and Thermodynamics of Systems with Long-Range
Interactions: Theory and Experiments}, edited by A. Campa, A. Giansanti, G. Morigi and F. Sylos Labini, AIP Conf. Proc. {\bf 965} 122 (2008)}
\bibitem{oxford}  {\small  {Long-Range Interacting Systems}, edited by T. Dauxois, S. Ruffo and L. Cugliandolo, Les Houches Summer School 2008, (Oxford: Oxford University Press, 2009)}
\bibitem{cdr}  {\small A. Campa, T. Dauxois, S. Ruffo,   Physics Reports {\bf 480}, 57 (2009)}
\bibitem{paddy}  {\small  T. Padmanabhan, Phys. Rep. {\bf 188}, 285 (1990)}
\bibitem{katz} {\small  J. Katz, Found. Phys. {\bf 33}, 223 (2003)}
\bibitem{ijmpb}  {\small P.H. Chavanis, Int. J. Mod. Phys. B {\bf 20}, 3113 (2006)}
\bibitem{angleaction}  {\small  P.H. Chavanis, Physica A {\bf 377}, 469 (2007)}
\bibitem{kindetail}  {\small  P.H. Chavanis, J. Stat. Mech P05019 (2010)}
\bibitem{heyvaerts}  {\small  J. Heyvaerts, Mon. Not. R. Astron. Soc. {\bf 407}, 355 (2010)}
\bibitem{newangleaction}  {\small  P.H. Chavanis, [arXiv:1107.1475]}
\bibitem{klimontovich}  {\small  Yu. L. Klimontovich, {The Statistical Theory of Non-Equilibrium Processes in a Plasma} (MIT press, Cambridge Massachusetts, 1967)}
\bibitem{kac} {\small  M. Kac, G.E. Uhlenbeck, P.C. Hemmer,  J. Math. Phys. {\bf 4}, 216 (1963)}
\bibitem{jeans}  {\small  J. Jeans, Mon. Not. R. Astron. Soc. {\bf 76}, 71 (1915)}
\bibitem{vlasov}  {\small  A.A. Vlasov, Zh. Eksp. Teor. Fiz. {\bf  8}, 291 (1938); J. Phys. (U.S.S.R.) {\bf  9}, 25 (1945)}
\bibitem{bh} {\small  W. Braun, K. Hepp,  Commun. Math. Phys. {\bf 56}, 101 (1977)}
\bibitem{lb}  {\small  D. Lynden-Bell, Mon. Not. R. Astron. Soc.  {\bf 136}, 101 (1967)}
\bibitem{miller}  {\small  J. Miller,   Phys. Rev. Lett.  {\bf 65}, 2137 (1990)}
\bibitem{rs}  {\small  R. Robert, J. Sommeria, J. Fluid Mech.  {\bf 229}, 291 (1991)}
\bibitem{incomplete} {\small  P.H. Chavanis, Physica A {\bf 365}, 102 (2006)}
\bibitem{ms} {\small  J. Messer, H. Spohn,  J. Stat. Phys. {\bf 29}, 561 (1982)}
\bibitem{bgm} {\small F. Bouchet, S. Gupta, D. Mukamel, Physica A {\bf 389}, 4389 (2010)}
\bibitem{pitaevskii}  {\small  E.M. Lifshitz, L.P.  Pitaevskii, {Physical Kinetics} (Pergamon Press, Oxford, 1981)}
\bibitem{lenard}  {\small A. Lenard, Ann. Phys. (N.Y.) {\bf 10}, 390 (1960)}
\bibitem{balescu2}  {\small   R. Balescu, Phys. Fluids {\bf 3}, 52 (1960)}
\bibitem{landau}  {\small L.D. Landau, Phys. Z. Sowj. Union  {\bf 10}, 154 (1936)}
\bibitem{kandrup1}  {\small H. Kandrup,  Astrophys. J. {\bf 244}, 316 (1981)}
\bibitem{hb3} {\small  P.H. Chavanis, Physica A {\bf 387}, 787 (2008)}
\bibitem{chandrat}  {\small S. Chandrasekhar, Astrophys. J.  {\bf 99}, 47 (1944) }
\bibitem{ps}  {\small I. Prigogine, G. Severne, Physica {\bf 32}, 1376 (1966) }
\bibitem{sh}  {\small G. Severne, M.J. Haggerty, Astrophys. Space. Sci. {\bf 45}, 287 (1976) }
\bibitem{hb2}  {\small P.H. Chavanis, Physica A  {\bf 361}, 81 (2006)}
\bibitem{spitzerbook}  {\small  L. Spitzer, {Dynamical Evolution of Globular Clusters} (Princeton Series in Astrophysics, Princeton, 1987)}
\bibitem{feix}  {\small O.C. Eldridge, M. Feix,  Phys. Fluids.  {\bf 6}, 398 (1962) }
\bibitem{kp}  {\small B.B. Kadomtsev, O.P. Pogutse, Phys. Rev. Lett.  {\bf 25}, 1155 (1970) }
\bibitem{bd}  {\small F. Bouchet, T.  Dauxois,  Phys. Rev. E {\bf 72}, 045103 (2005)}
\bibitem{cvb}  {\small  P.H. Chavanis, J. Vatteville, F.  Bouchet,  Eur. Phys. J. B {\bf  46}, 61 (2005)}
\bibitem{dawson}  {\small J. Dawson,  Phys. Fluids  {\bf 7}, 419 (1964)}
\bibitem{rouetfeix}  {\small J.L. Rouet, M. Feix,  Phys. Fluids B  {\bf 3}, 1830 (1991)}
\bibitem{campa}  {\small A. Campa, A.  Giansanti, G.  Morelli, Phys. Rev. E {\bf 76}, 041117 (2007) }
\bibitem{private}  {\small S. Gupta, S. Ruffo, private communication}
\bibitem{brucemiller} {\small B.N. Miller, Phys. Rev. E {\bf 53}, R4279 (1996)}
\bibitem{gouda} {\small T. Tsuchiya, N. Gouda, T. Konishi, Phys. Rev. E {\bf 53}, 2210 (1996)}
\bibitem{valageas2}  {\small P. Valageas, Phys. Rev. E {\bf 74}, 016606 (2006)}
\bibitem{joyce} {\small M. Joyce, T. Worrakitpoonpon, J. Stat. Mech. {\bf 10}, 12 (2010)}
\bibitem{ruffoN} {\small P. de Buyl, D. Mukamel, S. Ruffo, [arXiv:1012.2594]}
\bibitem{campachav}  {\small A. Campa, P.H. Chavanis, A.  Giansanti, G. Morelli,  Phys. Rev. E {\bf 78}, 040102 (2008) }
\bibitem{yamaguchi}  {\small Y. Yamaguchi, J.   Barr\'e, F. Bouchet, T.  Dauxois, S. Ruffo,   Physica A {\bf 337}, 36 (2004)}
\bibitem{collective}  {\small P.H. Chavanis, [arXiv:1107.1447]}    
\bibitem{hubbard1}  {\small J. Hubbard, Proc. R. Soc. Lond. {\bf 260}, 114 (1961a) }    
\bibitem{risken}  {\small H. Risken, The Fokker-Planck Equation (Springer, 1989)}    
\bibitem{hb4}  {\small  P.H. Chavanis, Physica A {\bf 387}, 1504 (2008)}    
\bibitem{kalnajs}  {\small  A.J. Kalnajs, Astrophys. Space. Sci. {\bf 13}, 279 (1971)} 
\bibitem{kandrup2}  {\small H. Kandrup,   Astro. Space. Sci.  {\bf 97}, 435 (1983)}   
\bibitem{chandra1}  {\small S. Chandrasekhar, Principles of Stellar Dynamics (University of Chicago Press, 1942)}
\bibitem{landaud}  {\small P.H. Chavanis, Eur. Phys. J. B {\bf 52}, 61 (2006)}
\bibitem{nardini}  {\small C. Nardini, S. Gupta, S. Ruffo, T. Dauxois, F. Bouchet, [arXiv:1111.6833]}
\bibitem{baldo1}  {\small F. Baldovin, P.H. Chavanis, E. Orlandini, Phys. Rev. E {\bf 79}, 011102 (2009)}
\bibitem{baldo2}  {\small P.H. Chavanis, F. Baldovin, E. Orlandini, Phys. Rev. E {\bf 83}, 040101(R) (2011)}
\bibitem{cd}  {\small P.H. Chavanis, L. Delfini, Eur. Phys. J. B {\bf 69}, 389 (2009)}
\bibitem{ichimaru}  {\small S. Ichimaru, Basic Principles of Plasma Physics (Westview, 1973)}







    
\end{thebibliography}
\end{document}